\documentclass[12pt,final]{article}
\usepackage[T1]{fontenc}
\usepackage{fourier}
\usepackage{calrsfs}
\DeclareMathAlphabet{\pazocal}{OMS}{zplm}{m}{n}
\usepackage{amssymb,amsthm,amsmath,amsfonts,enumitem,bbm,tikz,xr}
\usepackage{xcolor}
\usepackage{amsmath}
\usepackage{soul}
\usepackage{amsthm}
\usepackage{chngcntr}
\usepackage{apptools}
\AtAppendix{\counterwithin{lemma}{section}}
\AtAppendix{\counterwithin{prop}{section}}

\makeatletter
\def\thm@space@setup{%
  \thm@preskip=\parskip \thm@postskip=0pt
}
\makeatother
\makeatletter
\renewcommand\section{\@startsection {section}{1}{\z@}%
                                   {-1.5ex \@plus -1ex \@minus -.2ex}%
                                   {1.5ex \@plus.2ex}%
                                   {\normalfont\scshape}}
\renewcommand\subsection{\@startsection{subsection}{2}{\z@}%
                                     {-1.5ex\@plus -1ex \@minus -.2ex}%
                                     {0.25ex \@plus .2ex}%
                                     {\normalfont\itshape}}
\makeatother
\usepackage[multiple]{footmisc}
\usepackage[onehalfspacing]{setspace}
\usepackage{amsfonts}
\usepackage{graphicx}
\usepackage{subfig}
\usepackage{natbib}
\usepackage{clipboard}	
\newclipboard{PaperClipboard}	
\usepackage{xr-hyper} 
\usepackage[colorlinks,linkcolor=links,citecolor=cites,urlcolor=MyDarkBlue]{hyperref}

\usepackage[margin=1.25in]{geometry}
\usepackage{parskip}
\usepackage[justification=centering]{caption}
\usepackage{amssymb}
\usepackage{xcolor}
\usepackage{amsmath}
\usepackage{tikz}
\usepackage{pgfplots}
\usepackage{bbm}
\usepackage{xr}
\usepackage{nicefrac}
\usepackage[justification=centering]{caption}
\usepackage{amsthm}
\makeatletter
\def\thm@space@setup{%
  \thm@preskip=\parskip \thm@postskip=0pt
}
\makeatother
\usepackage{setspace}
\usepackage{amsfonts}
\usepackage{graphicx}
\usepackage{subfig}
\usepackage{natbib}
\usepackage[colorlinks,linkcolor=links,citecolor=cites,urlcolor=MyDarkBlue]{hyperref}

\usepackage{geometry}
\usepackage{parskip}
\usepackage{enumitem}
\usepackage[section]{placeins}

\definecolor{MyDarkBlue}{rgb}{0,0.08,0.45}
\definecolor{cites}{HTML}{324b13}
\definecolor{links}{HTML}{1a663b}
\definecolor{MyLightMagenta}{cmyk}{0.1,0.8,0,0.1}

\newtheorem{theorem}{Theorem}
\newtheorem{lemma}{Lemma} 
\newtheorem{prop}{Proposition}

\newtheorem{definition}{Definition} 
\newtheorem{corollary}{Corollary}

\newtheorem{rem}{Remark}

\newcommand{\dateindex}{\ensuremath{t}}
\newcommand{\timeindex}{\ensuremath{k}}

\newcommand{\transfer}{\ensuremath{x}}
\newcommand{\transferm}{\ensuremath{\transfer^\mechanism}}

\newcommand{\mechanismt}{\ensuremath{\mathbf{M}_\dateindex}}
\newcommand{\betat}{\ensuremath{\beta^{\mathbf{M}_\dateindex}}}
\newcommand{\alphat}{\ensuremath{\alpha^{\mathbf{M}_\dateindex}}}
\newcommand{\Mt}{\ensuremath{M^{\mathbf{M}_\dateindex}}}
\newcommand{\St}{\ensuremath{S^{\mathbf{M}_\dateindex}}}
\newcommand{\qt}{\ensuremath{q^{\mechanismt}}}
\newcommand{\transfert}{\ensuremath{\transfer^{\mechanismt}}}

\newcommand{\hagt}{\ensuremath{h_B^\dateindex}}

\newcommand{\mechanismtprm}{\ensuremath{\mathbf{M}_\dateindex^\prime}}

\newcommand{\betatstar}{\ensuremath{\beta^{\mathbf{M}_\dateindex^*}}}

\newcommand{\mechanism}{\ensuremath{\mathbf{M}}}
\newcommand{\Mm}{\ensuremath{M^\mechanism}}
\newcommand{\Sm}{\ensuremath{S^\mechanism}}
\newcommand{\betam}{\ensuremath{\beta^{\mechanism}}}

\newcommand{\qm}{\ensuremath{q^{\mechanism}}}
\newcommand{\tm}{\ensuremath{\transfer^{\mechanism}}}

\newcommand{\mechanisms}{\ensuremath{\pazocal{M}}}
\newcommand{\mechanismsc}{\ensuremath{\pazocal{M}_C}}
\newcommand{\mechanismc}{\ensuremath{\mathbf{M}}}

\newcommand{\mechanismcprm}{\ensuremath{\mathbf{M}^\prime}}
\newcommand{\betacprm}{\ensuremath{\beta^{\mechanismcprm}}}

\newcommand{\mechanismcstart}{\ensuremath{\mathbf{M}_\dateindex^*}}

\newcommand{\qcstart}{\ensuremath{q^{\mechanismcstart}}}
\newcommand{\tcstart}{\ensuremath{\transfer^{\mechanismcstart}}}

\newcommand{\prior}{\ensuremath{\mu_0}}

\newcommand{\eqbmsetprior}{\ensuremath{\pazocal{E}^*(\prior)}}

\newcommand{\posterior}{\ensuremath{\mu^\prime}}

\newcommand{\Posteriors}{\ensuremath{\Delta(V)}}
\newcommand{\game}{\ensuremath{G^\infty(}}
\newcommand{\Pbe}{\ensuremath{\langle\Gamma,(\pi_v,r_v)_{v\in V},\mu\rangle}}

\newcommand{\Pbestar}{\ensuremath{\langle\Gamma^*,(\pi_v^*,r_v^*)_{v\in V},\mu^*\rangle}}

\newcommand{\payoff}{\ensuremath{u}}
\newcommand{\payoffs}{\ensuremath{\payoff_S}}
\newcommand{\payoffh}{\ensuremath{\payoff_H}}
\newcommand{\payoffl}{\ensuremath{\payoff_L}}
\newcommand{\vl}{\ensuremath{v_L}}
\newcommand{\vh}{\ensuremath{v_H}}
\newcommand{\hatv}{\ensuremath{\hat{v}_L}}
\newcommand{\dv}{\ensuremath{\Delta v}}
\newcommand{\ml}{\ensuremath{m_L^*}}
\newcommand{\mh}{\ensuremath{m_H^*}}

\newcommand{\delay}{\ensuremath{\overline{\mu}}}
\newcommand{\delayone}{\ensuremath{\delay_1}}
\newcommand{\delayn}{\ensuremath{\delay_n}}
\newcommand{\delaynplus}{\ensuremath{\delay_{n+1}}}
\newcommand{\delaynminus}{\ensuremath{\delay_{n-1}}}

\newcommand{\eqpolicy}{\ensuremath{(\tau^*,q^*)}}
\newcommand{\Req}{\ensuremath{R^{\eqpolicy}}}

\newcommand{\joker}{\ensuremath{\mu^\star}}

\newcommand{\maxl}{\ensuremath{\overline{u}_L}}

\newcommand{\Types}{\ensuremath{V}}
\newcommand{\outcomes}{\ensuremath{\pazocal{O}^*}}
\newcommand{\allocations}{\ensuremath{A}}
\newcommand{\zpt}{\ensuremath{z_{(s_t,(q_t,\transfer_t))}(\mechanismt)}}
\newcommand{\znpt}{\ensuremath{z_{\emptyset}(\mechanismt)}}

\newcommand{\publict}{\ensuremath{h^t}}

\newcommand{\mechanismb}{\ensuremath{\mechanism^\prime}}

\newcommand{\mechanismbt}{\ensuremath{\mechanismb_t}}

\usetikzlibrary{positioning,chains,fit,shapes,calc,arrows,trees,decorations.pathmorphing,decorations.markings,hobby,calc,snakes,shapes.misc,patterns,backgrounds,quotes}
\usetikzlibrary{arrows}
\usetikzlibrary{trees,calc,snakes,fit,shapes,positioning}
\usetikzlibrary{decorations.pathmorphing}
\usetikzlibrary{decorations.markings}
\usetikzlibrary{decorations.shapes}
\tikzset{
  treenode/.style = {align=center, inner sep=0pt, text centered,
    font=\sffamily},
  arn_n/.style = {treenode, circle, black, font=\sffamily\bfseries, draw=black,
    fill=white, text width=1.5em},
  arn_r/.style = {treenode, circle, bg, draw=bg, 
    text width=1.5em},
  arn_x/.style = {treenode, circle, orange,font=\sffamily\bfseries,
   text width=2em}
}
\tikzset{
  invisible/.style={opacity=0},
  visible on/.style={alt={#1{}{invisible}}},
  alt/.code args={<#1>#2#3}{%
    \alt<#1>{\pgfkeysalso{#2}}{\pgfkeysalso{#3}} 
  },
}
\tikzstyle{every picture}+=[remember picture]

\tikzset{
    photon/.style={decorate, decoration={snake}, draw=red}}
\tikzset{electron/.style={draw=blue, postaction={decorate},decoration={markings,mark=at position .55 with {\arrow[draw=blue]{>}}}}}
\tikzset{electron2/.style={draw=red,very thick, postaction={decorate},decoration={markings,mark=at position .55 with {\arrow[draw=red,very thick]{>}}}}}
\tikzset{gluon/.style={->,thick,decorate, draw= mLightBrown,
        decoration={coil,amplitude=4pt, segment length=5pt}}}
 \tikzset{gluon2/.style={thick,decorate, draw=mLightBrown,
        decoration={coil,amplitude=4pt, segment length=5pt}}}
\tikzset{nero/.style={decorate,draw=black}}
\tikzset{bianco/.style={decorate,draw=bg}}
 \tikzset{ every node/.style={inner sep=0pt,minimum size=1mm},
  nsnode/.style={draw,circle,black},
  nnnode/.style={draw,circle,black,fill=black},
  asnode/.style={draw,circle,myblue,fill=myblue},
  bsnode/.style={draw,circle,blue,fill=blue},
  csnode/.style={draw,circle,red,fill=red, minimum size=2mm},
  every fit/.style={inner sep=-1.5pt,text width=1cm}  }

\newcommand{\belief}{\ensuremath{\mu}}

\newcommand{\policynull}{\ensuremath{(\tau_0,q_0)}}
\newcommand{\policynullb}{\ensuremath{(\tau_0^\prime,q_0^\prime)}}

\newcommand{\minseller}{\ensuremath{\underline{\payoff}_S}}

\newcommand{\measurablem}{\ensuremath{U^\prime}}
\newcommand{\measurablea}{\ensuremath{A^\prime}}

\newcommand{\Payoffs}{\ensuremath{U_S}}

\newcommand{\payoffhl}{\ensuremath{\payoff_{H|L}}}
\newcommand{\eqbmset}{\ensuremath{\pazocal{E}^*}}
\newcommand{\participate}{\ensuremath{\hat{\pi}}}
\newcommand{\postedpayoff}{\ensuremath{\payoffs^*}}
\newcommand{\virtual}{\ensuremath{VS}}
\newcommand{\maxvirtual}{\ensuremath{\overline{\virtual}}}
\newcommand{\maxpayoff}{\ensuremath{\overline{\payoff}}}

\newcommand{\mechanismzb}{\ensuremath{\mechanism_0^\prime}}
\newcommand{\qzb}{\ensuremath{q^{\mechanismzb}}}
\newcommand{\betazb}{\ensuremath{\beta^{\mechanismzb}}}
\newcommand{\tzb}{\ensuremath{\transfer^{\mechanismzb}}}

\newcommand{\continuation}{\ensuremath{(\payoffh,\payoffs)}}
\newcommand{\continuationb}{\ensuremath{(\payoffh^\prime,\payoffs^\prime)}}

\newcommand{\policyprior}{\ensuremath{(\tau_{\prior}^*,q_{\prior}^*)}}

\newcommand{\postedpayoffh}{\ensuremath{\payoffh^*}}

\newcommand{\cdf}{\ensuremath{G}}

\newcommand{\Delayn}{\ensuremath{[\delayn,\delaynplus)}}

\newcommand{\payoffb}{\ensuremath{\payoff^\prime}}
\newcommand{\varphim}{\ensuremath{\varphi^\mechanism}}

\usepackage{xr}
\title{Optimal mechanism for the sale of a durable good\thanks{
We thank the Editor Simon Board and three anonymous referees for excellent comments which substantially improved the paper. We would like to thank Rahul Deb, Frederic Koessler, Dan Quigley, Pablo Schenone, and especially Max Stinchcombe, as well as audiences at Cowles, SITE, and Stony Brook,  for thought-provoking questions and illuminating discussions. Vasiliki Skreta is grateful for generous financial support through the ERC consolidator grant 682417 ``Frontiers in design." This research was supported by grants from the National Science Foundation (SES-1851744 and SES-1851729).}}
\author{Laura Doval\thanks{Columbia University and CEPR. E-mail: \href{mailto:laura.doval@columbia.edu}{\texttt{laura.doval@columbia.edu}}}\and Vasiliki Skreta\thanks{University of Texas at Austin, University College London, and CEPR. E-mail: \href{mailto:vskreta@gmail.com}{\texttt{vskreta@gmail.com}}}}

\begin{document}
\maketitle
\begin{abstract}
A buyer wishes to purchase a durable good from a seller who in each period chooses a mechanism under limited commitment. The buyer's valuation is binary and fully persistent. We show that posted prices implement all equilibrium outcomes of an infinite-horizon, mechanism selection game. Despite being able to choose mechanisms, the seller can do no better and no worse than if he chose prices in each period, so that he is subject to Coase's conjecture. Our analysis marries insights from information and mechanism design with those from the literature on durable goods. We do so by relying on the revelation principle in \cite{doval2020mechanism}.

\end{abstract}

\newpage
\section{Introduction}\label{sec:intro}
We characterize the equilibrium outcomes of an infinite-horizon, mechanism-selection game between a durable-good seller and a privately informed buyer under \emph{limited commitment}, so that the seller can commit to today's mechanism, but not to the mechanism he will offer if no sale occurs. \autoref{theorem:characterization} shows that all equilibrium outcomes can be implemented via posted prices. We construct a Perfect Bayesian equilibrium of the mechanism-selection game, which achieves the seller's unique equilibrium payoff, and we show that it implements the \emph{essentially} unique equilibrium outcome.\footnote{Whenever the equilibrium outcome is not unique, all equilibrium outcomes are achieved also via a sequence of posted prices.} In this equilibrium, as long as a sale has not occurred, the seller will choose a mechanism that can be implemented as a posted price. Despite being able to choose from a rich set of mechanisms, the seller can do no better and no worse than if he could only choose prices in each period.
\newline\indent
In our game, an uninformed seller faces a privately informed buyer, whose valuation is binary, fully persistent, and strictly above the seller's marginal cost.\footnote{In the \emph{no gap} case, all equilibrium outcomes can still be implemented in posted prices, but equilibrium multiplicity arises as in \cite{ausubel1989reputation} (see \autoref{fnt:vl-0}).} In each period, as long as the good has not been sold, the seller offers the buyer a \emph{mechanism}, the rules of which determine the allocation for that period. A mechanism consists of (i) a set of \emph{input} messages for the buyer, and (ii) for each input message, a distribution over \emph{output} messages and allocations. While the seller observes the output message and the allocation, he does not observe the input message the buyer submits to the mechanism. Thus, when designing the mechanism, the seller gets to design \emph{how much} he observes about the buyer's choices and hence, design his beliefs about the buyer's value. The combination of mechanism design and information design elements is key to our characterization.
\newline\indent
Our analysis bridges the literatures on mechanism design and on the durable-good monopolist, especially the work of \citet*{gul1986foundations}. To see this, it is useful to review the two main steps involved in the proof of \autoref{theorem:characterization}. First, we construct an assessment that is identical along the path to that in \cite{hart1988contract}, which we dub the \emph{posted-prices assessment}. In this assessment, along the path of play, the seller sells the good using a decreasing sequence of prices, which reflect that conditional on the good not being sold the seller assigns less probability to the buyer's valuation being high. Using the property that beliefs go down in the posted-prices assessment, we show that the seller's payoff in this assessment is a lower bound on the seller's equilibrium payoff. The logic of the argument is analogous to the argument in \cite*{gul1986foundations} that there is a unique equilibrium in the gap case.
\newline\indent
Second, we argue that the seller's payoff under the posted-prices assessment is an upper bound on the seller's equilibrium payoff. To do so, we rely on an \emph{auxiliary} program, that only involves the seller (see \ref{eq:opt} in \autoref{sec:new-opt}). In this program, the seller maximizes the dynamic analogue of the \emph{virtual surplus}, by choosing a Bayes' plausible distribution over posteriors and for each posterior (i) a probability of trade and (ii) a vector of equilibrium continuation payoffs. 
We arrive at the program defined in \ref{eq:opt} by relying on the tools in our previous work, \cite{doval2020mechanism}. The main theorem in \cite{doval2020mechanism} allows us to simplify the class of mechanisms the seller will offer in any equilibrium of the game and the buyer's equilibrium behavior. This step reduces the search of the optimal sequence of mechanisms to those that satisfy, loosely speaking, a sequence of \emph{participation} and \emph{truthtelling} constraints, allowing us for the most part to ignore the buyer as a player. Like in standard mechanism design, the participation constraint of the low-valuation buyer and the truthtelling constraint of the high-valuation buyer impose an upper bound on the revenue the seller can extract within a period (\autoref{sec:new-opt}). Replacing this upper bound in the seller's payoff provides us with a dynamic analogue of the virtual surplus (\autoref{eq:virtual}), where the seller's payoff is written as a function of the allocation, but also the continuation payoffs. 
\newline\indent
\label{relaxed-intro}We show that the value of the program defined in \ref{eq:opt} coincides with the seller's payoff in the posted-prices assessment and hence, that the seller cannot do better than in the posted-prices equilibrium. 
Note that the solution to the auxiliary program corresponds to the solution to the \emph{relaxed} program in mechanism design, where only a subset of the incentive constraints have been imposed. However, as argued by \cite{laffont1993theory} and more recently by \cite{breig2020repeated}, it is not immediate that the solution to the relaxed problem satisfies \emph{all} constraints and hence, can be implemented as an equilibrium outcome. Nevertheless, as we discuss in the conclusions, we expect that in settings with transferable utility, the study of the analogous problem to \ref{eq:opt} provides a natural benchmark to understand the properties of the principal's optimal mechanism.
\newline\indent
The significance of our results is two-fold. 
First, to the best of our knowledge, this is the first paper to characterize optimal mechanisms under limited commitment and persistent private information in an infinite-horizon setting. Such characterization has proved elusive. Indeed, the set of tools available to tackle the difficulties with the revelation principle under limited commitment do not readily apply to infinite-horizon settings (see, for instance, the seminal work of \cite{bester2001contracting,bester2007contracting}, and the discussion in the related literature). 
In \cite{doval2020mechanism}, we introduce a tool akin to the revelation principle for mechanism-selection games under limited commitment that applies to a broad class of games, including infinite-horizon ones. It is the application of this tool that allows us to argue that the mechanism that we characterize is the optimal one amongst \emph{all} mechanisms the seller could have offered the buyer under limited commitment.
\newline\indent
Second, while the optimality of posted-prices is evocative of \cite{skreta2006sequentially}, there are two reasons why this result should not be taken for granted. First, our model is not an infinite-horizon version of the one studied in \cite{skreta2006sequentially}, since we allow the seller to choose from a class of mechanisms that includes those in \cite{skreta2006sequentially}. In \cite{doval2020mechanism}, we study a two-period version of the model in \cite{skreta2006sequentially}, but we allow the seller to offer mechanisms like those in this paper. We show that when the seller is sufficiently patient it is not an equilibrium for the seller to post a price in each period. It follows that we cannot take limits using the equilibrium outcome in \cite{skreta2006sequentially} to analyze the equilibrium outcomes of the game we study, even after showing that the game ends in finite time. Second, \cite{breig2020repeated} shows in a binary-value model with a perishable good that posted-prices may not be optimal: Indeed, the seller may benefit from using random delivery contracts.

\paragraph{Related Literature:} The paper contributes mainly to three strands of literature. The first strand, similar to this paper, derives optimal mechanisms when the designer has limited commitment. Most papers in that literature examine either finite-horizon settings (see \citealp{laffont1988dynamics,skreta2006sequentially,skreta2015optimal,deb2015dynamic,fiocco2015consumer,beccuti2018dynamic}), or infinite-horizon settings, imposing restrictions on the class of contracts that can be offered (e.g., \citealp{gerardi2020dynamic}), or on the solution concept (e.g., \citealp{acharya2017progressive}).\footnote{\cite{beccuti2018dynamic} lie somewhat in between these two strands because they take limits of their finite-horizon results to draw conclusions about the infinite-horizon setting. However, they do not show that this limit corresponds to the seller's revenue-maximizing equilibrium in the infinite-horizon game.} Underlying these restrictions is that the results in both \cite{bester2001contracting} and \cite{skreta2006sequentially} do not readily extend to infinite-horizon settings. For instance, the result in \cite{bester2001contracting} applies only if the principal is earning his highest payoff consistent with the agent's payoff (see Lemma 1 in \citealp{bester2001contracting}). Thus, implicit in their multistage extension is a restriction to equilibria of the mechanism-selection game that possess a Markov structure, which, as shown by \cite{ausubel1989reputation}, may not be enough to characterize the principal's best equilibrium payoff. The approach in \cite{skreta2006sequentially} has the advantage that the set of incentive feasible outcomes has a well-understood structure. It is not clear, however, how to incorporate the principal's sequential rationality constraints in infinite-horizon settings in a tractable way.
The second strand is the literature that follows the observation in \cite{coase1972durability} that the durable-good monopolist faces a time-inconsistency problem, which in turn limits his monopoly power. The papers in the durable-good monopolist literature (\citealp*{stokey1981rational,bulow1982durable,gul1986foundations,sobel1991durable,ortner2017durable}) study price dynamics
and establish (under some conditions) Coase's conjecture.\footnote{Other papers, like \cite{wolinsky1991durable}, \cite{mcafee2008capacity} and \cite{board2014outside} study variations on Coase's original problem and their implications for Coase's conjecture.} Related to this literature are the problems of dynamic bargaining with one-sided incomplete information\footnote{Recently, \cite{peski2019alternating} studies alternating bargaining games where players can offer menus.} (e.g., \citealp*{sobel1983multistage,fudenberg1985infinite,ausubel1989reputation}). In all these papers, the seller's inability to commit reduces monopoly profits. 

Finally, as will become clear from the analysis, the paper contributes to the literature on information design (\cite*{aumann1995repeated} and \cite{kamenica2011bayesian}), highlighting its potential to provide tractable characterizations of equilibrium outcomes in games. 
Contrary to most of these papers, however, the seller aims to persuade his future self, as opposed to another player, highlighting the role of information as a commitment device (see, e.g., \cite{carrillo2000strategic}, and recently, \cite{habibi2020motivation}).

\paragraph{Organization} The rest of the paper is organized as follows. \autoref{sec:model} describes the model; \autoref{sec:rp} summarizes the results in \cite{doval2020mechanism} used to simplify the analysis that follows. Section \ref{sec:main} presents the main result of the paper, \autoref{theorem:characterization}. \autoref{sec:new-opt} introduces the auxiliary program \ref{eq:opt} and studies its properties. \autoref{sec:proof} reviews the main steps of the proof of \autoref{theorem:characterization}.  \autoref{sec:conclusions} concludes. All proofs not in the main text are in the Appendix.

\section{Model}\label{sec:model}
\textbf{Primitives:} Two players, a seller and a buyer, interact over infinitely many periods. The seller owns one unit of a durable good to which he attaches value $0$. The buyer has private information: before her interaction with the seller starts, she observes her valuation $v\in\{\vl,\vh\}\equiv V$, with $0<\vl<\vh$. Let $\dv\equiv\vh-\vl$ denote the difference in values. Let \prior\ denote the probability that the buyer's valuation is \vh\ at the beginning of the game. In what follows, we denote by \Posteriors\ the set of distributions on $V$.
\newline\indent
An allocation in period $\dateindex$ is a pair $(q,\transfer)\in\{0,1\}\times\mathbb{R}$, where $q$ indicates whether the good is traded ($q=1$) or not ($q=0$), and \transfer\ is a payment from the buyer to the seller. Let $A$ denote the set of allocations. The game ends the first time the good is traded.\footnote{While the set of allocations is $\{0,1\}\times\mathbb{R}$, we allow the seller to offer randomizations on \allocations\ and hence induce \emph{fractional} assignments of the good.} 
\newline\indent
Payoffs are as follows. If in period $\dateindex$, the allocation is $(q,x)$, the flow payoffs are $u_B(q,x,v)=vq-x$ and $u_S(q,x)=x$ for the buyer and the seller, respectively. The seller and the buyer maximize the expected discounted sum of flow payoffs. They share a common discount factor $\delta\in(0,1)$.

\textbf{Mechanisms:} To introduce the timing of the game, we first define the action space of the seller.  In each period, the seller offers the buyer a mechanism. Following \cite{myerson1982optimal}, \cite{bester2007contracting}, and \cite{doval2020mechanism}, we define a mechanism as follows. A mechanism, $\mechanism=(\Mm,\Sm,\varphi^{\mechanism})$, consists of a set of \emph{input} messages \Mm, a set of \emph{output} messages \Sm, and a transition probability $\varphi^{\mechanism}$ from $\Mm$ to $\Sm\times A$.\footnote{Throughout, we assume that \Mm\ and \Sm\ are Polish spaces, that is, they are separable, completely metrizable topological spaces. Note that the set of allocations $A$ is also a Polish space, and therefore $\Sm\times A$ is a Polish space. For a Polish space $X$, let $\Delta(X)$ denote the set of Borel measures on $X$. We endow $\Delta(X)$ with the weak$^*$ topology. Thus, $\Delta(X)$ is also a Polish space (\citealp{aliprantis2013infinite}). For any two measurable spaces $X$ and $Y$, a mapping $\zeta:X\mapsto\Delta(Y)$ is a \emph{transition probability from $X$ to $Y$ } if for any measurable $C\subseteq Y$, $\zeta(C|x)$ is a measurable real valued function of $x\in X$.} For instance, \Mm\ can be the set of buyer values, $V$, and \Sm\ be the set of seller beliefs about the buyer's value, \Posteriors. In this case, the mechanism associates to each report a distribution over beliefs and allocations. We endow the seller with a collection $(M_i,S_i)_{i\in\pazocal{I}}$ of input and output messages in which each $M_i$ contains at least two elements, and each $S_i$ contains $\Posteriors$.\footnote{We show in \cite{doval2020mechanism} that it is without loss of generality to take \Mm\ to be finite when the set of types is finite.\label{fnt:cardinality}} Denote by $\mechanisms$
the set of all mechanisms with message sets $(M_i,S_i)_{i\in\pazocal{I}}$.\footnote{We restrict the seller to choose mechanisms with input and output messages in $(M_i,S_i)_{i\in\pazocal{I}}$
to have a well-defined action space for the seller. This allows us to have a well-defined set of deviations, avoiding set-theoretic issues related to self-referential sets. The analysis in \cite{doval2020mechanism} shows that the choice of the collection plays no further role in the analysis.}

\textbf{Timing:} In each period \dateindex, as long as the good has not been sold, the game proceeds as follows. First, the seller offers the buyer a mechanism, \mechanism. Observing the mechanism, the buyer decides whether to participate or not. If she does not participate in the mechanism, the good is not sold and no payments are made. If she instead chooses to participate, she sends a message $m\in\Mm$, which is unobserved by the seller. An output message and an allocation, $(s,q,\transfer)$, are drawn from $\varphi^\mechanism(\cdot|m)$. The output message and the allocation are observed by both the seller and the buyer. If the good is not traded, the game proceeds to period $\dateindex+1$.
\newline\indent The above defines an infinite-horizon dynamic game, which we denote by $G_{\mechanisms}^\infty(\prior)$ to remind the reader that the seller's beliefs at the beginning of the game are given by $\prior$. Public histories in this game are
\begin{align*}
h^\dateindex=(\mechanism_0,\pi_0,s_0,(q_0,\transfer_0),\dots,\mechanism_{\dateindex-1},\pi_{\dateindex-1},s_{\dateindex-1},(q_{\dateindex-1},\transfer_{\dateindex-1})),
\end{align*}
where $\pi_\timeindex\in\{0,1\}$ denotes the buyer's participation decision, with the restriction that if $\pi_\timeindex=0$, then $s_\timeindex=\emptyset$, i.e., no output message is generated, and $(q_\timeindex,\transfer_\timeindex)=(0,0)$.\footnote{While there is no output message when the buyer does not participate in the mechanism, we denote this by $s=\emptyset$ to keep 
the length of all histories the same.} Let $H^\dateindex$ denote the set of all period $t$ public histories; they capture what the seller knows through period $t$. 
\newline\indent
A \emph{buyer} history consists of the public history of the game together with the buyer's reports into the mechanism and her private information.
%
Formally, a buyer history is an element
\begin{align*}
\hagt=(\mechanism_0,\pi_0,m_0,s_0,(q_0,\transfer_0),\dots,\mechanism_{\dateindex-1},\pi_{\dateindex-1},m_{\dateindex-1},s_{\dateindex-1},(q_{\dateindex-1},\transfer_{\dateindex-1})).
\end{align*}
Let $H_B^\dateindex$ denote the set of buyer histories through period $\dateindex$. The buyer also knows her valuation and hence, when her valuation is $v$, a history through period \dateindex\ is an element of $\{v\}\times H_B^\dateindex$. 
\newline\indent\textbf{Strategies and beliefs:} Since the seller's action space, \mechanisms, is an uncountable set of functions, we model the seller's behavioral strategy $(\Gamma_t)_{t=0}^{\infty}$ following \cite{aumann1964mixed}. That is, endow $[0,1]$ with the Borel $\sigma$-algebra and the Lebesgue measure. Then, $\Gamma_t$ is defined as a jointly measurable function from $H^t\times[0,1]$ to $\mechanisms$.\footnote{As we explain in \autoref{appendix:pbe}, the set \mechanisms\ is a Polish space; we endow it with its Borel $\sigma$-algebra.} We denote the collection $(\Gamma_t)_{t=0}^{\infty}$ by $\Gamma$. The buyer's participation strategy is given by $\pi_{t}:\Types\times H_B^t\times\mechanisms\mapsto[0,1]$. Conditional on participating in the mechanism, \mechanismt, her reporting strategy is a transition probability from $\Types\times H_B^t\times\mechanisms\times\{1\}$ to $\cup_{i\in\pazocal{I}}M_i$ such that $r_{t}(h_B^t,\mechanismt)\in\Delta(\Mt)$. We denote the tuple $(\pi_t(v,\cdot),r_t(v,\cdot))$ by $(\pi_{tv},r_{tv})$ and the collection $(\pi_{tv},r_{tv})_{t=0}^{\infty}$ by $(\pi_v,r_v)$.
\newline\indent
A belief for the seller at the beginning of time $\dateindex$, history $h^\dateindex$, is a distribution $\mu_t(h^\dateindex)\in\Delta(V\times H_B^\dateindex(h^\dateindex))$, where $H_B^\dateindex(h^\dateindex)$ is the set of buyer histories consistent with the public history, $h^\dateindex$. The belief system, $(\mu_t)_{t=0}^\infty$, is denoted by $\mu$.

\textbf{Solution concept:} We are interested in studying the Perfect Bayesian equilibrium (henceforth, PBE) payoffs of this game, where PBE is defined \emph{informally} as follows. An assessment, $\Pbe$, is a PBE if the following hold:
\begin{enumerate}
\item $\Pbe$ satisfies sequential rationality, and
\item $\mu$ satisfies Bayes' rule where possible.
\end{enumerate}
\autoref{appendix:pbe} contains the formal statement. For now, we note that if the principal's strategy space were finite and the mechanisms used by the principal had finite support, then this coincides with the definition in \cite{fudenberg1991perfect}.\footnote{The only difference between Bayes' rule where possible and consistency in sequential equilibrium is the following. Under PBE, the principal can assign zero probability to a type and then, after the agent deviates, can assign positive probability to that same type.}

The prior \prior\ together with the strategy profile $(\Gamma,(\pi_v,r_v)_{v\in V})$ induce a distribution over the \emph{terminal nodes} $V\times H_B^\infty$. We are interested instead on the distribution it induces over the payoff-relevant outcomes, $V\times A^\infty$. We say that a distribution $\eta\in\Delta(V\times A^{\infty})$ is a PBE outcome if there is a PBE assessment \Pbe\ which induces $\eta$. We denote by $\pazocal{O}_{\mechanisms}^*(\prior)$ the set of PBE outcomes and by $\pazocal{E}_{\mechanisms}^*(\prior)\subseteq\mathbb{R}^3$ the set of PBE payoffs of $G_{\mechanisms}^{\infty}(\prior)$. We denote a generic element of $\pazocal{E}_{\mechanisms}^*(\prior)$ by $\payoff\equiv(u_L,u_H,u_S)$, where $u_S$ is the seller's payoff and $u_L,u_H$ denote the buyer's payoff when her value is $\vl,\vh$, respectively. 

%
\autoref{theorem:characterization} characterizes $\pazocal{O}_{\mechanisms}^*(\prior)$ and $\eqbmset_{\mechanisms}(\prior)$. In particular, we show that the \emph{essentially} unique equilibrium outcome can be achieved via a sequence of posted prices so that the seller of a durable good can do no better and no worse than by using posted prices.
\subsection{Revelation Principle}\label{sec:rp}
There are at least two reasons that the game $G_{\mechanisms}^{\infty}(\prior)$ is not simple to analyze. First, the seller's action space is large and, a priori, it is not clear which
mechanisms could be ruled out from consideration. Second, fixing a seller's strategy, and hence
a sequence of mechanisms faced by the buyer, we still need to understand the buyer's best response in the game induced by the sequence of mechanisms. 

 Let $\game\prior)$ denote the same game in the previous section, except that in each period (i) both the buyer and the seller observe a draw from a public randomization device $\omega\sim U[0,1]$, and (ii) the seller's action space is the set of \emph{canonical mechanisms} denoted by \mechanismsc\ and defined as follows. \mechanismsc\ is the set of all mechanisms where the set of input and output messages are the buyer's values and the seller's beliefs about the buyer's value, respectively. That is, $(M,S)=(V,\Posteriors)$.
%
Let $\outcomes(\prior)$ denote the set of PBE outcomes and  \eqbmsetprior\ denote the set of PBE payoffs of $\game\prior)$.

\autoref{lemma:rp} summarizes the key implications of \cite{doval2020mechanism} for our analysis, which we explain below:
%

\begin{lemma}[\citealp{doval2020mechanism}]\label{lemma:rp}
$G_{\mechanisms}^\infty(\prior)$ and $\game\prior)$ have the same set of equilibrium outcomes, i.e., $\pazocal{O}_{\mechanisms}^*(\prior)=\pazocal{O}^*(\prior)$. 
%

Moreover, let $\eta\in\outcomes(\prior)$. Then, there exists a PBE assessment \Pbe\ of \space$\game\prior)$ that induces $\eta$ and satisfies the following properties:
\begin{enumerate}
\item\label{itm:rp-1} For all histories $h^\dateindex$, the buyer participates in the mechanism offered by the seller at that history and truthfully reports her type, with probability $1$,
\item\label{itm:rp-2} For all histories $h^\dateindex$, if the mechanism offered by the seller at $h^\dateindex$ outputs posterior \posterior,  the seller's updated equilibrium beliefs about the buyer coincide with \posterior,
\item\label{itm:rp-4} For all histories \publict, the mechanism offered by the seller at \publict, $\varphi^{\mechanismt}:V\mapsto\Delta(\Posteriors\times A)$ can be decomposed into a \emph{communication device} $\betat:V\mapsto\Delta\Posteriors$ and an \emph{allocation rule} $\alphat:\Posteriors\mapsto\Delta(A)$,\footnote{That is, for all measurable subsets $\measurablem\subset\Posteriors$ and $\measurablea\subset\allocations$, we have that for all $v\in V$,
\[\varphi^{\mechanismt}(\measurablem\times\measurablea|v)=\int_{\measurablem}\alphat(\measurablea|\belief)\betat(d\belief|v).\]}
\item\label{itm:rp-3} The strategy of the buyer depends only on her private valuation and the public history.
\end{enumerate}
\end{lemma}
\autoref{lemma:rp} has several implications. 
%
Part \ref{itm:rp-1} of \autoref{lemma:rp} implies the mechanisms chosen by the seller in equilibrium must satisfy a \emph{participation} constraint and an \emph{incentive compatibility constraint} for each buyer value and each public history. As in the case of commitment to long-term mechanisms, part \ref{itm:rp-1} simplifies the analysis of the buyer's behavior, by reducing it to a series of constraints (see Equations \ref{eq:participation-t} and \ref{eq:truthtelling-t}).

Part \ref{itm:rp-2} implies that the mechanism's output message encodes all of the information that the seller has in equilibrium about the buyer's value. In particular, conditional on observing the output message, the allocation carries no more information about the buyer's value. As a consequence, conditional on the output message, the allocation can be drawn independently of the buyer's report. This, in turn, delivers the decomposition of $\varphi^{\mechanismt}$ described in part \ref{itm:rp-4}. 

Part \ref{itm:rp-4} implies that the choice of mechanism at history \publict\ can be equivalently thought of as the choice of an experiment (the communication device, \betat) and an allocation rule (the \alphat). The decomposition of $\varphi^{\mechanismt}$ allows us to separately optimize on the allocation \emph{given} a particular communication device, and then optimize on the communication device. As in the literature on information design, it is convenient to work with the distribution over posteriors induced by the experiment \betat, which we denote by $\tau^{\mechanismt}$ and is defined as follows. For all Borel subsets $U^\prime\subseteq[0,1]$,\label{bayes-rp}
\begin{align}\label{eq:bayes-rp}\tag{BC$_{\mu_t(h^\dateindex)}$}
\int_{U^\prime}\tau^{\mechanismt}(d\belief_{t+1})=\sum_{v\in V}\belief_t(\publict)(v)\betat(U^\prime|v),
\end{align}
where $\belief_t(\publict)$ is the seller's belief about the buyer's value at \publict. Furthermore, as in the literature on mechanism design with quasilinear utilities, we can write $\alphat(\cdot|\belief)$ as an expected payment, $\transfert(\belief)$, and a probability of trade, $\qt(\belief)$. 

Part \ref{itm:rp-3} implies the set of PBE payoffs of $\game\prior)$ coincides with the set of \emph{Public PBE} payoffs of $\game\prior)$ (\citealp{athey2008collusion}), allowing us to invoke the results in \cite{abreu1990toward} and \cite{athey2008collusion} to argue the assessment we construct in the next section is indeed a PBE assessment.

%
%
The rest of the paper studies the equilibrium outcomes and payoffs of $\game\prior)$ and when we refer to a PBE assessment, we mean one that satisfies the conditions of \autoref{lemma:rp}.
\begin{rem}In what follows, we abuse notation in the following two ways. First, because valuations are binary, we can think of an element in $\Delta(V)$ (a distribution over $\vl$ and $\vh$) as an element of the interval $[0,1]$ (the probability assigned to $\vh$). We use the latter formulation in what follows. That is, whereas the mechanism outputs a distribution over $\vl$ and $\vh$, we index this distribution by the probability of $\vh$. Second, even though $\beta(\cdot|v)$ is a measure over $\Posteriors$ (in this case a c.d.f.), we sometimes write $\beta(\belief|v)$ when $\beta$ has an atom at \belief. \end{rem}

\section{Main result}\label{sec:main}

\autoref{sec:main} contains the main result of the paper: \autoref{theorem:characterization} characterizes the equilibrium outcomes and payoffs of \game\prior). To state \autoref{theorem:characterization}, we proceed as follows. First, we describe a PBE assessment \Pbestar, which is singled out by the proof of \autoref{theorem:characterization}. Second, we  explain why the equilibrium outcome induced by \Pbestar\ can be implemented via a sequence of posted prices. Finally, we state \autoref{theorem:characterization}.
%
%
%
\paragraph{An assessment in posted prices:}\label{assessment} The proof of \autoref{theorem:characterization} singles out a particular PBE assessment, \Pbestar, that implements the seller's unique equilibrium payoff, which we now define informally (the formal statement is in \autoref{appendix:sellerfull}). 

The assessment \Pbestar\ is essentially the same along the path of play to those constructed by \cite{fudenberg1985infinite} and \cite{hart1988contract}. The assessment is defined by an increasing sequence of threshold beliefs $\delay_0=0<\delayone<\dots<\delayn<\dots$ such that if the seller's belief is in $[\delayn, \delaynplus)$, then it takes $n$ periods for the good to be sold to \vl, at which point the game ends. The number of periods before \vl\ buys the good determines both the rents for \vh\ and the seller's revenue. 

To understand how the sequence of thresholds is determined, note that if a seller with belief \prior\ sells the good at a price of \vl, his revenue can be written as follows:
\label{hatv}
\begin{align}\label{eq:hatv}
\vl=\prior(\vh-\dv)+(1-\prior)\vl=\prior\vh+(1-\prior)\underbrace{\left(\vl-\frac{\prior}{1-\prior}\dv\right)}_{\equiv \hatv(\prior)}.
\end{align}
The first equality represents revenue as the surplus extracted from each type. The second equality represents revenue as the \emph{virtual surplus}, where the value of allocating the good to $\vl$ is adjusted to capture that when $\vl$ is served, so is $\vh$, which leaves rents to $\vh$. The threshold $\delayone\equiv\nicefrac{\vl}{\vh}$ is the belief at which the virtual value, $\hatv(\prior)$, equals $0$. At that belief, the seller is indifferent between serving the buyer for both of her values and excluding the low-valuation buyer.

For a seller with belief $\prior<\delayone$, \vl\ is the maximum revenue that the seller can make. Instead, a seller with belief $\prior>\delayone$ prefers to take one period to trade with \vl\ to trading immediately with \vl. To see this, note that if the good is sold with probability $1$ in the next period, the price at that point is \vl. Thus, the maximum price that the seller can sell the good for in the first period is $\vh-\delta(\vh-\vl)=\vl+(1-\delta)\dv$. It is easy to check that when $\prior>\delayone$, the seller prefers to take one period to sell the good to \vl. When the seller's prior belief is \delayone\ the seller is indifferent between selling the good to \vl\ in one period and selling the good to \vl\ immediately. Indeed, the subscript $1$ in \delayone\ reflects that the seller takes at most one period to trade with \vl. Recursively, one can define \delayn\ as the belief at which the seller is indifferent between trading with \vl\ in $n$ period or $n-1$ periods.

In the assessment, play proceeds as follows. If the seller's prior is such that  $\prior\in [\delayn,\delaynplus)$, then the seller chooses a mechanism that induces two beliefs, $1$ and \delaynminus. At belief $1$, the good is sold and the transfer is $\vl+(1-\delta^n)\dv$, whereas at belief \delaynminus, the good is not sold and the transfer is $0$ (see \autoref{fig:optimal-prior}). Subsequently, a seller with belief $\delay_m$ for $m\leq n-1$, chooses a mechanism that induces two beliefs, $1$ (with allocation $(1,\vl+(1-\delta^m)\dv)$) and $\delay_{m-1}$ (with allocation (0,0)) (see \autoref{fig:optimal-cutoff}). Thus, starting from \prior, the game ends in $n$ periods. \autoref{fig:belief-dynamics} illustrates how beliefs fall conditional on no trade.
%
\begin{figure}[h!]
\begin{minipage}{\textwidth}
\begin{center}
\subfloat[Mechanism for $\prior\in\Delayn$ under \Pbestar]{\begin{tikzpicture}[scale=1.5,thick]
\node(origin) at (0,0){};
\node[label=below:{}](end) at (10,0){};
\draw[|->](origin)--(end);

\node[label=below:{$\delayone$}](d1) at (1,0){};
\draw[|-](d1)--(end);

\node[label=below:{$\delay_2$}](d2) at (2,0){};
\draw[|-](d2)--(end);

\node[label=below:{\dots}](dots) at (3,0){};
\node[label=below:{$\delaynminus$}](dnminus) at (4,0){};
\draw[|-](dnminus)--(end);

\node[label=below:{$\delayn$}](dn) at (5,0){};
\draw[|-](dn)--(end);

\node[label=below:{\textcolor{blue}{$\prior$}}](prior) at (6,-0.1){};
\draw[|-](6,0)--(end);

\node[label=below:{$\delaynplus$}](dn+1) at (7,0){};
\draw[|-](dn+1)--(end);
\node[label=below:{\dots}](dots) at (8,0){};

\node[label=below:{1}](one) at (9.5,0){};
\draw[|-](one)--(end);
\path[every node/.style={font=\sffamily\small},->]
    (6,0) edge[bend left] node [left] {} (one);
    \path[every node/.style={font=\sffamily\small},->]
    (6,0) edge[bend right] node [left] {} (dnminus);
 
\draw[->](4,-0.4)--(4,-0.8);
\node[label=below:{$q_{\prior}^*(\delaynminus)=0$}]at (4,-0.8){};
\draw[->](9.5,-0.4)--(9.5,-0.8);
\node[label=below:{$q_{\prior}^*(1)=1$}]at (9.5,-0.8){};
\end{tikzpicture}
\label{fig:optimal-prior}}
\end{center}
\end{minipage}
\begin{minipage}{\textwidth}
\begin{center}
\subfloat[Mechanism for $\prior=\delaynminus$ under \Pbestar]{\begin{tikzpicture}[scale=1.5,thick]
\node(origin) at (0,0){};
\node[label=below:{}](end) at (10,0){};
\draw[|->](origin)--(end);

\node[label=below:{$\delayone$}](d1) at (1,0){};
\draw[|-](d1)--(end);

\node[label=below:{$\delay_2$}](d2) at (2,0){};
\draw[|-](d2)--(end);

\node[label=below:{$\delay_{n-2}$}](dots) at (3,0){};
\draw[|-](dots)--(end);
\node[label=below:{$\delaynminus$}](dnminus) at (4,0){};
\draw[|-](dnminus)--(end);

\node[label=below:{$\delayn$}](dn) at (5,0){};
\draw[|-](dn)--(end);

\node[label=below:{\textcolor{blue}{$\prior$}}](prior) at (6,-0.1){};
\draw[|-](6,0)--(end);

\node[label=below:{$\delaynplus$}](dn+1) at (7,0){};
\draw[|-](dn+1)--(end);
\node[label=below:{\dots}](dots2) at (8,0){};

\node[label=below:{1}](one) at (9.5,0){};
\draw[|-](one)--(end);
\path[every node/.style={font=\sffamily\small},->]
    (dnminus) edge[bend left] node [left] {} (one);
    \path[every node/.style={font=\sffamily\small},->]
    (dnminus) edge[bend right] node [left] {} (dots);
 
\draw[->](3,-0.4)--(3,-0.8);
\node[label=below:{$q_{\delaynminus}^*(\delay_{n-2})=0$}]at (3,-0.8){};
\draw[->](9.5,-0.4)--(9.5,-0.8);
\node[label=below:{$q_{\delaynminus}^*(1)=1$}]at (9.5,-0.8){};
\end{tikzpicture}
\label{fig:optimal-cutoff}}
\end{center}
\end{minipage}
\begin{minipage}{\textwidth}
\begin{center}
\subfloat[Belief dynamics starting from \prior.]{\begin{tikzpicture}[scale=1.5,thick]
\node(origin) at (0,0){};
\node[label=below:{}](end) at (10,0){};
\draw[|->](origin)--(end);

\node[label=below:{$\delayone$}](d1) at (1,0){};
\draw[|-](d1)--(end);

\node[label=below:{$\delay_2$}](d2) at (2,0){};
\draw[|-](d2)--(end);

\node[label=below:{\dots}](dots) at (3,0){};
\node[label=below:{$\delaynminus$}](dnminus) at (4,0){};
\draw[|-](dnminus)--(end);

\node[label=below:{\textcolor{blue}{$\prior$}}](prior) at (6,-0.1){};
\draw[|-](6,0)--(end);

\node[label=below:{$\delayn$}](dn) at (5,0){};
\draw[|-](dn)--(end);

\node[label=below:{$\delaynplus$}](dn+1) at (7,0){};
\draw[|-](dn+1)--(end);
\node[label=below:{\dots}](dots2) at (8,0){};

\node[label=below:{1}](one) at (9.5,0){};
\draw[|-](one)--(end);
    \path[every node/.style={font=\sffamily\small},->]
    (6,0) edge[bend right] node [left] {} (dnminus); 
    \path[every node/.style={font=\sffamily\small},->]
    (dnminus) edge[bend right] node [left] {} (dots);
    \path[every node/.style={font=\sffamily\small},->]
    (dots) edge[bend right] node [left] {} (d2);
\path[every node/.style={font=\sffamily\small},->]
    (d2) edge[bend right] node [left] {} (d1);
    \path[every node/.style={font=\sffamily\small},->]
    (d1) edge[bend right] node [left] {} (origin);           
\end{tikzpicture}\label{fig:belief-dynamics}}
\end{center}
\end{minipage}
\caption{Mechanism and belief dynamics under \Pbestar\ as a function of \prior.}
\end{figure}

More precisely, the PBE assessment, \Pbestar, is as follows:

\begin{enumerate}[leftmargin=*]
\item Along the equilibrium path:
\begin{enumerate}
\item If at history $h^t$, the seller's beliefs, $\mu^*(h^t)$, are in $[\delay_0,\delay_1)$, he chooses a mechanism such that $(\qcstart(\mu^*(h^t)),\tcstart(\mu^*(h^t)))=(1,\vl)$ and the communication device satisfies that $\betatstar(\mu^*(h^t)|v)=1$ for $v\in\{\vl,\vh\}$.
 \item If at history $h^t$, the seller's beliefs, $\mu^*(h^t)$, are in $[\delayn,\delaynplus)$ for $n\geq1$, the seller's mechanism satisfies the following. First, it induces two posteriors, $\delaynminus$ and $1$. Second, the allocation rule satisfies that $(\qcstart(1),\tcstart(1))=(1,\vl+(1-\delta^n)\dv)$, whereas $(\qcstart(\delaynminus),\tcstart(\delaynminus))=(0,0)$. Finally, the communication device maps $\vl$ to $\delaynminus$, whereas it maps $\vh$ to both $\delaynminus$ and $1$ with positive probability. The probabilities $\betatstar(\delaynminus|\vh),\betatstar(1|\vh)$ are chosen so that when the seller observes $\delaynminus$, his updated belief coincides with \delaynminus, that is
 \begin{align*}
 \delaynminus=\frac{\belief^*(\publict)\betatstar(\delaynminus|\vh)}{\belief^*(\publict)\betatstar(\delaynminus|\vh)+(1-\belief^*(\publict))}
 \end{align*}
\end{enumerate}
\item Off the equilibrium path, the seller's strategy coincides with the above, except that when $\mu^*(h^t)=\delayn$ for some $n\geq 1$, the seller may randomize between the mechanism he offers on the path of play when his belief is $\delayn$ and the one he offers on the path of play when his belief is $\delaynminus$.\footnote{The need for mixing arises for technical reasons: it ensures that the buyer's continuation payoffs
when her valuation is \vh\ are upper-semicontinuous and, thus, guarantees that a best response exists after any deviation by the seller (see \autoref{appendix:complete}). Indeed, we appeal to the results in \cite{simon1990discontinuous} to simultaneously determine the buyer's best response and the seller's mixing.}
 \item At each history $h^t$, the buyer's best response to the seller's equilibrium offer at $h^t$ is to participate in the mechanism and truthfully report her valuation.
\end{enumerate}

\paragraph{An implementation in posted prices:}\label{implementation} We now argue that there is an implementation in posted prices of the equilibrium outcome induced by \Pbestar.
Clearly, when the seller's beliefs are below $\delayone$, the seller's mechanism corresponds to selling the good at a price of $\vl$. Consider then the case in which the seller's beliefs are in $[\delayone,\delay_2)$. Note that when the buyer's valuation is \vh\ and the realized allocation is trade, then her payoff is $\vh-\vl-(1-\delta)\dv=\delta\dv$. On the other hand, when the buyer's valuation is \vh\ and the realized allocation is no trade, then the seller's beliefs next period are $\delay_0=0$, so that the buyer's continuation payoff is $\delta\dv$. That is, the buyer with valuation \vh\ is indifferent between obtaining the good at price $\vl+(1-\delta)\dv$ and not obtaining the good, and paying a price of \vl\ in the next period. Since the buyer with valuation is \vh\ is indifferent between these two options, she is willing to mix between buying at price $\vl+(1-\delta)\dv$ and not obtaining the good. She does so in a way that the seller's belief is $\delay_0$ when the allocation is $(0,0)$. Since $\delay_0=0$, it implies that when the seller's belief is in $[\delayone,\delay_2)$, the buyer buys with probability $1$ at a price of $\vl+(1-\delta)\dv$. Working recursively through the equations, one can show that when the seller's prior is in $[\delayn,\delaynplus)$, the mechanism is equivalent to posting a price of $\vl+(1-\delta^n)\dv$. In this case, the low-valuation buyer chooses the $(0,0)$ allocation, whereas the high-valuation buyer mixes so that the seller's belief is $\delaynminus$ when the allocation is $(0,0)$.
\begin{rem}[Direct vs. indirect implementation]
It is interesting to contrast the implementation under the PBE assessment \Pbestar\ described above with the implementation via posted prices. In the former, the buyer is truthful and the seller \emph{rations} the high-valuation buyer, which slows down the rate at which the seller's beliefs fall conditional on the good not being sold. Instead, in the implementation via posted prices, the high-valuation buyer misreports her type with positive probability, which, like rationing, prevents the seller from becoming pessimistic too quickly about the buyer's valuation. Since both implementations are payoff equivalent, the seller cannot do better with rationing than with posted prices. 
\end{rem}
\label{dg}
\begin{rem}[Rationing]\label{rem:dg}
The mechanism used by the seller in \Pbestar\ is what  \cite{denicolo1999rationing} dub rationing. In a two-period model, \cite{denicolo1999rationing} show that if the seller observes only whether trade happens, the seller may prefer to ration high valuation buyers instead of posting a price in the first period, in order to induce a strong demand in the second period. Whereas rationing does not dominate posted prices when values are binary, we show in \cite{doval2020mechanism}  that both posted prices and rationing as in \cite{denicolo1999rationing} may be dominated when values are drawn from a continuum in a two-period model.
%
\end{rem}

\autoref{theorem:characterization} states that there is a unique equilibrium payoff for the seller and it coincides with the payoff that the seller obtains under \Pbestar. Furthermore, the low-valuation buyer's payoff is also unique and equal to $0$ (see \autoref{lemma:vl-0}). Finally, except at the threshold beliefs $\{\delayn\}_{n\geq 1}$, the high-valuation buyer's payoff is also unique. The multiplicity of \vh's payoffs arises because when the seller's prior belief is \delayn\ for $n\geq1$, the seller is indifferent between trading with \vl\ in $n$ periods and in $n-1$ periods, whereas \vh's rents are higher when \vl\ receives the good in $n-1$ periods. All equilibrium payoff vectors for \vh\ can be obtained by randomizing between trading with \vl\ in $n$ and $n-1$ periods.%
%
\footnote{When the seller's \emph{prior} belief coincides with $\delayn$ for $n\geq1$, there is another PBE assessment which is payoff equivalent to \Pbestar\ for the seller, but delivers a payoff of $\postedpayoffh(\prior)/\delta$ to the high-valuation buyer. For $n\geq 2$, this PBE assessment is identical to \Pbestar\ except that in the first period the seller offers a mechanism that induces two posteriors, $\delay_{n-2}$ and $1$. When the posterior is $1$, the allocation is $(1,\vl+(1-\delta^{n-1})\dv)$, whereas when the posterior is $\delay_{n-2}$, the allocation is $(0,0)$.
Instead, when $n=1$, the seller sells the good with probability $1$ at a price of $\vl$. Verifying that this is also a PBE assessment follows immediately from the arguments in \autoref{appendix:equilibrium}.}

To state \autoref{theorem:characterization}, let $\postedpayoffh(\prior)$ and $\postedpayoff(\prior)$ denote the high-valuation buyer and seller's payoff under the assessment \Pbestar\ when the seller's prior belief is \prior.
\begin{theorem}\label{theorem:characterization}
\Pbestar\ is a PBE assessment. Furthermore, the set of equilibrium payoffs of \game\prior) is the following. For $\prior\in[0,\delay_1)$, $\eqbmsetprior=(0,\postedpayoffh(\prior),\postedpayoff(\prior))$. Furthermore, for all $n\geq 1$, and all $\prior\in[\delayn,\delaynplus)$
\begin{align}\label{eq:delay-n}
\eqbmsetprior=\left\{\begin{array}{cc}(0,\postedpayoffh(\prior),\postedpayoff(\prior))&\text{if }\prior>\delayn\\
\{0\}\times[\postedpayoffh(\prior),\nicefrac{\postedpayoffh(\prior)}{\delta}]\times\{\postedpayoff(\prior)\}&\text{otherwise}
\end{array}\right..
\end{align}
\end{theorem}
\autoref{theorem:characterization} together with the above discussion implies:
\begin{corollary}\label{corollary:posted-prices-implementation}
All equilibrium outcomes of \game\prior) can be implemented via a sequence of posted prices.\footnote{In the gap case, that is, when $\vl=0$, multiplicity of the equilibrium outcomes and payoffs obtains as in \cite{ausubel1989reputation} and yet, equilibrium outcomes can also be sustained using posted prices. For instance, the seller can obtain the monopoly profit $\prior\vh$ by setting a price of \vh\ in period $1$ and thereafter, not selling the good.\label{fnt:vl-0}}
\end{corollary}
\autoref{sec:proof} reviews the main steps of the proof of \autoref{theorem:characterization}. \autoref{sec:uniqueness} shows that \postedpayoff(\prior) is a lower bound and an upper bound on the seller's equilibrium payoff. \autoref{sec:pbe-assessment} describes the arguments to show that \Pbestar\ is a PBE assessment. To show that \postedpayoff(\prior) is an upper bound on the seller's equilibrium payoff and that \Pbestar\ is a PBE assessment, we rely on an auxiliary program, which we denote by \ref{eq:opt}, and define in \autoref{sec:new-opt}, to which we turn next. 
\section{Auxiliary program: maximization of the virtual surplus}\label{sec:new-opt}\label{opt}
We now introduce a program, denoted by  \ref{eq:opt}, that is our main tool of analysis. We use it in \autoref{sec:uniqueness} to show both that the seller cannot do better than in the posted-prices equilibrium and that \Pbestar\ is a PBE assessment. In this program, the seller maximizes 
the dynamic analogue of the virtual surplus, which we denote by \virtual. Recall that in standard mechanism design the virtual surplus of a mechanism is an upper bound on the revenue from the mechanism that only depends on the probability of trade, since the transfers are determined by the buyer's participation and truthtelling constraints (\autoref{sec:preliminaries} provides the details of this derivation in our setting). As we explain below, \virtual\ depends on the distribution over posteriors induced by the mechanism's communication device (recall \autoref{eq:bayes-rp} on page \pageref{bayes-rp}). Moreover, because of the dynamic nature of our problem, it also depends on the continuation payoffs. 

%
%
Formally, 
\label{opt}
\begin{align}\label{eq:opt}\tag{OPT}
&\max_{\policynull,\continuation}\virtual(\policynull,\continuation,\prior),\\
&\text{s.t.}\left\{\begin{array}{l}\tau_0\in\Delta(\Posteriors)\text{ is Bayes' plausible for \prior\ }\\
q_0:\Posteriors\mapsto[0,1]\\
(\forall\belief_1\in\Posteriors)(0,\payoffh(\belief_1),\payoffs(\belief_1))\in\eqbmset(\belief_1)\end{array}\right.\nonumber,
\end{align}
%
where
\begin{align}\label{eq:virtual}
&\virtual(\policynull,(\payoffh,\payoffs),\prior)\equiv\int_{\Posteriors}\left[\begin{array}{c}q_0(\belief_{1})(\belief_{1}\vh+(1-\belief_{1})\hatv(\prior))+\\
(1-q_0(\belief_1))\delta\left(\payoffs(\belief_1)+(1-\belief_1)\left(\frac{\belief_1}{1-\belief_1}-\frac{\prior}{1-\prior}\right)\payoffh(\belief_1)\right)\end{array}\right]\tau_0(d\belief_{1}),
\end{align}
defines the \emph{virtual surplus}. Note that the virtual surplus only depends on a mechanism's probability of trade, $q_0$, and distribution over posteriors, $\tau_0$. In what follows, we identify a mechanism \mechanism\ with its induced $(\tau^{\mechanism},\qm)$ and refer to the latter also as the mechanism.

In \ref{eq:opt}, we allow the seller with prior \prior\ not only to choose his most preferred mechanism \policynull, (accruing its virtual surplus), but also his preferred continuation payoffs, subject to the constraint that these continuation payoffs are actually equilibrium payoffs. That is, $(0,\payoffh(\belief_1),\payoffs(\belief_1))\in\eqbmset(\belief_1)$. The notation anticipates the result in \autoref{lemma:vl-0} below that the low-valuation buyer's payoff is $0$ in \emph{any} equilibrium (see \autoref{sec:preliminaries}).

The objective in \ref{eq:opt}, which is the virtual surplus, consists of two terms. The first term\label{virtual-surplus-terms-page}
\[q_0(\belief_{1})\left(\belief_{1}\vh+(1-\belief_{1})\hatv(\prior)\right),\]
represents how much surplus the seller can extract subject to the rents he must leave to the high-valuation buyer (recall \autoref{eq:hatv}). Indeed, whenever $\belief_{1}\neq 1$ and $q_0(\belief_{1})>0$, the seller sells the good with positive probability to the low-valuation buyer. In that case, \vh\ gets rents equal to $q_0(\belief_{1})\dv$, which the seller pays with probability $\prior$. This, in turn, explains why the virtual value of \vl, $\hatv(\cdot)$ is evaluated at the seller's prior belief, $\prior$, instead of the posterior belief, $\belief_{1}$. The second term 
\begin{align*}
(1-q_0(\belief_1))\delta\left(\payoffs(\belief_1)+(1-\belief_1)\left(\frac{\belief_1}{1-\belief_1}-\frac{\prior}{1-\prior}\right)\payoffh(\belief_1)\right),
\end{align*}
accounts for the rents the \vh\ receives in terms of continuation payoffs, which limit how much surplus the seller can extract in period $0$ whenever he induces no trade with positive probability. The term pre-multiplying \vh's continuation payoffs, $\payoffh(\belief_1)$, reflects that conditional on no trade, the seller's beliefs about the buyer's value may change so that
 the optimal mechanism from period $1$ onward may not coincide with what is optimal from period $0$ onward. The \emph{wedge} between the likelihood ratio of $\belief_{1}$ and \prior\ reflects this disagreement and it is the source of the time inconsistency of the commitment solution.\label{simon-time-consistent}
%
%
 
While the construction of the upper bound on the seller's payoff is reminiscent of standard observations in mechanism design, it is important to note one challenge relative to this literature, which explains why we allow the seller to select continuation payoffs in \ref{eq:opt}. The standard argument for why a revenue-maximizing seller can secure the virtual surplus of a mechanism is that if he were not, then there would be a way to increase transfers so that the buyer's participation and truthtelling constraints would still be satisfied, whereas the seller's revenue would increase.\label{feasibility-upper-bound} This argument does not translate immediately  to a game. Because continuation play may depend on the mechanism \mechanism\ chosen by the seller, it may not be possible for him to secure the mechanism's virtual surplus. To do so, the seller would need to offer a mechanism $\mechanismb$ that coincides with \mechanism, except that transfers are higher. However, even if \mechanismb\ induces the same distribution over allocations and seller's beliefs as \mechanism, this could trigger lower continuation payoffs for the seller, preventing him from obtaining the virtual surplus of \mechanism. The program defined in \ref{eq:opt} allows us to circumvent this difficulty: Because the seller can choose both the mechanism \emph{and} the continuation payoffs in \ref{eq:opt}, the seller does not need to consider how his choice of mechanism may adversely affect his continuation payoffs.
%
%
%

\autoref{sec:preliminaries} contains the details of the construction of the virtual surplus defined in \autoref{eq:virtual} and verifies that it is an upper bound on the seller's equilibrium payoff (\autoref{lemma:vs-upper bound}). The reader interested in the properties of the solution to \ref{eq:opt} and how it is used in the proof of \autoref{theorem:characterization} can proceed to \autoref{sec:opt} with little loss of continuity.

\subsection{Derivation of the virtual surplus}\label{sec:preliminaries}
We now derive the virtual surplus defined in \autoref{eq:virtual} by relying on \autoref{lemma:rp} and show that it is an upper bound on the seller's equilibrium payoff. To do so, consider a PBE assessment, \Pbe. Fix a history \publict\ and let \mechanismt\ denote the mechanism offered by the seller at \publict\ under the assessment. Let $(\tau^{\mechanismt},\qt)$ denote the distribution over posteriors and the probability of trade associated with \mechanismt. Furthermore, the PBE assessment specifies 
continuation payoffs $(\payoffh^{\mechanismt},\payoffs^{\mechanismt})$ when the seller offers mechanism \mechanismt\ at history \publict. In what follows, we show that the seller's equilibrium payoff at \publict\ satisfies the following:
\begin{lemma}\label{lemma:vs-upper bound}
The seller's payoff at \publict, \Payoffs(\publict) satisfies
\begin{align*}
\Payoffs(\publict)\leq\virtual((\tau^{\mechanismt},\qt),(\payoffh^{\mechanismt},\payoffs^{\mechanismt}),\belief_t(\publict)).
\end{align*}
\end{lemma}


To see why \autoref{lemma:vs-upper bound} holds, note that \autoref{lemma:rp} implies that the seller's equilibrium payoff at \publict, $U_S(\publict)$, can be written as:
\begin{align}\label{eq:revenue-t}
U_S(\publict)=\int_{\Posteriors}\left(\transfert(\belief_{t+1})+(1-\qt(\belief_{t+1}))\payoffs^{\mechanismt}(\belief_{t+1})\right)\tau^{\mechanismt}(d\belief_{t+1}),
\end{align}
where $\payoffs^{\mechanismt}(\belief_{t+1})$ is short-hand notation for the continuation payoff of the seller when at history \publict, he offers \mechanismt\ and the output message is $\belief_{t+1}$.\footnote{This continuation payoff can also depend on \publict, but we omit this dependence to simplify notation.} \autoref{eq:revenue-t} uses \autoref{lemma:rp} as follows. First, the seller's payoff from offering \mechanismt\ is written under the assumption that the buyer participates in the mechanism and truthfully reports her value. Second, it uses \autoref{lemma:rp} to write the mechanism in terms of the distribution over posteriors $\tau^{\mechanismt}$ and the allocation $(\qt,\transfert)$. 

In particular, the mechanism \mechanismt\ together with the continuation payoffs $(\payoffl^{\mechanismt},\payoffh^{\mechanismt})$ satisfy the following constraints. First, the buyer prefers to participate in the mechanism for both her values, that is for $v\in\{\vl,\vh\}$ the following holds:
\small
\begin{align}\label{eq:participation-t}\tag{PC$_{\publict,v}$}
\int_{\Posteriors}\left(v\qt(\belief_{t+1})-\transfert(\belief_{t+1})+(1-\qt(\belief_{t+1}))\delta\payoff_v^{\mechanismt}(\belief_{t+1})\right)\betat(d\belief_{t+1}|v)\geq\payoff_v^{\mechanismt}(\emptyset), 
\end{align}
\normalsize
where $\payoff_v^{\mechanismt}(\emptyset)$ is short-hand notation for the continuation payoff of the buyer when at history \publict, the seller offers \mechanismt\ and the buyer rejects. Also, the buyer prefers to truthfully report her value to the mechanism, that is, for $v\in\{\vl,\vh\}$ and $v^\prime\neq v$, the following holds:
\small
\begin{align}\label{eq:truthtelling-t}\int_{\Posteriors}\left(v\qt(\belief_{t+1})-\transfert(\belief_{t+1})+(1-\qt(\belief_{t+1}))\delta\payoff_v^{\mechanismt}(\belief_{t+1})\right)(\betat(d\belief_{t+1}|v)-\betat(d\belief_{t+1}|v^\prime))\geq0.\tag{IC$_{h^t,v,v^\prime}$}
\end{align}
\normalsize
There is one more way in which the above expressions implicitly use \autoref{lemma:rp}. By \autoref{lemma:rp}, the assessment \Pbe\ is a Public PBE, so that the continuation payoff vector $\payoff^{\mechanismt}(\belief_{t+1})\equiv(\payoffl^{\mechanismt}(\belief_{t+1}),\payoffh^{\mechanismt}(\belief_{t+1}),\payoffs^{\mechanismt}(\belief_{t+1}))$ is an equilibrium payoff vector of $\game\belief_{t+1})$. Formally, $\payoff^{\mechanismt}(\belief_{t+1})\in\eqbmset(\belief_{t+1})$.
%
%

Equations \ref{eq:participation-t} and \ref{eq:truthtelling-t} are analogous to the participation and incentive compatibility constraints one would obtain in mechanism design except for the following: The participation constraint is potentially type-dependent (the right hand side is $\payoff_v^{\mechanismt}(\emptyset)$). To obtain a participation constraint that more closely resembles that of the commitment case, we show that the low-valuation buyer's equilibrium payoff is $0$ in \emph{any} equilibrium of $\game\prior)$:
\begin{lemma}\label{lemma:vl-0}
For all $\prior\in\Posteriors$, if $\payoff\in\eqbmsetprior$, then $\payoffl=0$. 
\end{lemma}
\autoref{lemma:vl-0} implies that when the buyer's value is \vl, her utility from participation is $0$. Thus, we can write \autoref{eq:participation-t} when $v=\vl$ as follows:
\small
\begin{align}\label{eq:participation-t-vl}
\int_{\Posteriors}\left(\vl\qt(\belief_{t+1})-\transfert(\belief_{t+1})\right)\betat(d\belief_{t+1}|\vl)=0,
\end{align}
\normalsize
where to obtain \autoref{eq:participation-t-vl}, we have used \autoref{lemma:vl-0} twice in \autoref{eq:participation-t}. First, to replace the continuation payoffs by $0$ both after participation and after rejection of \mechanismt. Second, to change the inequality in \autoref{eq:participation-t} to an equality.  After all, the buyer's utility  when she participates in the mechanism \mechanismt\ is her equilibrium payoff at \publict, and hence an element of $\eqbmset(\belief_t(\publict))$.
\autoref{eq:participation-t-vl} can be used to rewrite \autoref{eq:truthtelling-t} for $v=\vh$ as follows:
\begin{align}\label{eq:truthtelling-t-vh}
&\int_{\Posteriors}\left(\vh\qt(\belief_{t+1})-\transfert(\belief_{t+1})+(1-\qt(\belief_{t+1}))\delta\payoffh^{\mechanismt}(\belief_{t+1}))\right)\betat(d\belief_{t+1}|\vh)\geq\nonumber\\
&\int_{\Posteriors}\left(\dv\qt(\belief_{t+1})+(1-\qt(\belief_{t+1}))\delta\payoffh^{\mechanismt}(\belief_{t+1}))\right)\betat(d\belief_{t+1}|\vl).
\end{align}

Equations \ref{eq:participation-t-vl} and \ref{eq:truthtelling-t-vh} already impose constraints on the maximum revenue the seller can make in period $t$. Indeed, like in standard mechanism design, the participation constraint for \vl\ and the truthtelling constraint for \vh\ determine the maximum expected payment the seller can extract from the buyer in period $t$. Replacing the upper bound on the expected payments obtained from these equations in the seller's payoff (\autoref{eq:revenue-t}), we obtain an upper bound on the seller's revenue at \publict\ when he offers mechanism \mechanismt. It is immediate to check that this upper bound corresponds to  $\virtual((\tau^{\mechanismt},\qt),(\payoffh^{\mechanismt},\payoffs^{\mechanismt}),\belief(\publict))$. \autoref{lemma:vs-upper bound} then follows.

Having established why the optimization program in \ref{eq:opt} is an upper bound on the seller's equilibrium payoff, \autoref{sec:opt} studies properties of its solution which are important for the proof of \autoref{theorem:characterization}.

%

\subsection{Properties of the solution to \ref{eq:opt}}\label{sec:opt}
The main result in this section, \autoref{prop:new-b1}, shows two properties of the solution to the program defined in \ref{eq:opt} that are important to prove that the seller's maximum equilibrium payoff is \postedpayoff(\prior). As we explain next, a key difficulty that the program in \ref{eq:opt} allows us to circumvent is the analysis of the \emph{belief dynamics} in the game.

An important result in the bargaining literature is the \emph{skimming lemma}, which states that incentive compatibility of the buyer's behavior implies that the expected discounted probability of trade of \vh\  is higher than that of \vl\ (see, for instance, \citealp{fudenberg1985infinite}). This property immediately implies that along the path of play the seller's beliefs fall conditional on no trade. As a consequence, prices must fall along the path of play. 

Contrast this to the game we analyze in which the seller offers mechanisms that enable him to design how much he observes about the buyer's choices. In particular, the seller can choose how fast he learns about the buyer's value conditional on no trade; he could even choose to become more optimistic about the buyer's value conditional on no trade.\footnote{While the truthtelling equations \ref{eq:truthtelling-t} can be used to derive a ``monotonicity'' condition analogous to that in the skimming lemma, it only implies that on average the expected probability of trade of \vh\ must be higher than that of \vl:
\small
\begin{align*}
\int_{\Posteriors}\left(\dv\qt(\belief_{t+1})+(1-\qt(\belief_{t+1}))\payoffh^{\mechanismt}(\belief_{t+1})\right)\left(\betat(d\belief_{t+1}|\vh)-\betat(d\belief_{t+1}|\vl)\right)\geq0.
\end{align*}} 
On the one hand, this would allow the seller to avoid the belief dynamics associated with posted-prices and hence, avoid the temptation to trade more often with \vl\ in future rounds. On the other hand, this comes at a cost. Recall that \autoref{lemma:rp} implies that the mechanism's allocation is measurable with respect to the information generated by the mechanism and this information is, in turn, subject to the Bayes' plausibility constraint. This implies that for the seller's beliefs conditional on no trade to fall slowly (or not fall at all), it must be that the seller is selling the good to \vh\ with small probability.  

%
%
%
It turns out that the program defined in \ref{eq:opt} is useful to discipline belief dynamics. While it may not be obvious how to rule out that there is an equilibrium in which the seller's beliefs may sometimes go up conditional on no trade, it turns out that this is never the case in a solution to \ref{eq:opt}. Indeed, as we establish in \autoref{prop:new-b1} below, it is never optimal to not sell the good and induce a belief above the prior. Furthermore, whenever $\prior>\delayone$, conditional on selling the good with positive probability, the seller sells the good only to the high-valuation buyer. \label{belief-dynamics-solution}
\begin{prop}\label{prop:new-b1}
Suppose that $\prior\geq\delayone$ and
let \policynull,\continuation\ denote a solution to \ref{eq:opt}. Then, the following hold:
\begin{enumerate}
\item\label{itm:no-delay-above} It is never optimal to induce a belief $\belief_1\geq\prior$ and not sell the good. That is, letting $A=[\prior,1)$, then
\begin{align*}
\int_A(1-q_0(\belief_1))\tau_0(d\belief_1)=0.
\end{align*}
\item\label{itm:trade-1} Furthermore, if $\prior>\delayone$ and the seller induces $\belief_1$ and sells the good (that is, $q_0(\belief_1)>0$), then $\belief_1=1$.
\end{enumerate}
\end{prop}
The proof is in \autoref{appendix:new-b1}. In what follows, we provide intuition for \autoref{prop:new-b1}, starting from part \ref{itm:no-delay-above}. To see why it is not optimal to not sell the good while at the same time inducing a belief $\belief_1\geq\prior$, note the following. First, associated to any continuation payoff, $(\payoffh(\belief_1),\payoffs(\belief_1))$, there is a mechanism chosen by the seller when his belief is $\belief_1$, and continuation payoffs for the seller and the buyer in the event that the good is not sold. This implies that, conditional on inducing a belief $\belief_1$, the seller with belief \prior\ could always choose today the mechanism and the continuation payoffs associated with $(\payoffh(\belief_1),\payoffs(\belief_1))$ in a solution to \ref{eq:opt}.  Second, the seller with belief \prior\ below $\belief_1$ pays rents to \vh\ with lower probability than the seller with belief $\belief_1$ (the term pre-multiplying \payoffh\ in \autoref{eq:virtual} is positive). It follows that the seller with belief \prior\ prefers to accrue today the payoff from the mechanism (and continuation payoffs) that induce $(\payoffh(\belief_1),\payoffs(\belief_1))$, contradicting that it is optimal to induce $\belief_1$ and not sell the good. 
%
%
%
%
%

That part \ref{itm:trade-1} holds follows from the observation that a seller with prior above \delayone\ prefers to trade with \vl\ in at least two periods (recall that $\hatv(\prior)<0$ when $\prior>\delayone$). Thus, it is never optimal to sell the good to \vl\ with positive probability today. It follows that if $q_0(\belief_1)>0$, then the seller must assign the good to \vh, and hence $\belief_1=1$.

\autoref{prop:new-b1} implies that a solution to \ref{eq:opt} never induces posteriors in $[\prior,1)$. This, in turn, delivers the following expression for the value of \ref{eq:opt}, where \virtual(\prior) denotes the value function of the program in \ref{eq:opt}:
\begin{align}\label{eq:opt-1}
\virtual(\prior)\equiv\max_{\tau_0,\continuation}&\tau_0(\{1\})\vh+\int_{[0,\prior)}\delta\left(\payoffs(\belief_1)+\left(\frac{\belief_1}{1-\belief_1}-\frac{\prior}{1-\prior}\right)\payoffh(\belief_1)\right)\tau_0(d\belief_1).
\end{align}
\normalsize
\autoref{eq:opt-1} simply states that the solution to \ref{eq:opt} can be described by the probability of selling to \vh\ today (the probability of inducing a belief $\belief_1=1$) and the probability with which the good is not sold and a belief below the prior is induced. One distribution over posteriors is of particular interest in what follows: the one that splits \prior\ between $1$ (with $q_0(1)=1$) and $\belief_1<\prior$ (with $q_0(\belief_1)=0$). Bayes' plausibility implies that the weights on $1$ and $\belief_1$ are
\begin{align*}
\frac{\prior-\belief_1}{1-\belief_1} \quad\text{ and }\quad\frac{1-\prior}{1-\belief_1},
\end{align*}
respectively. Note that this is precisely the type of mechanism that the seller uses in the posted-prices assessment. \autoref{corollary:indifference} shows that these are essentially the distributions over posteriors that solve the problem in \autoref{eq:opt-1}:
\begin{corollary}\label{corollary:indifference}
The value of the program in \autoref{eq:opt-1} equals the value of 
\small
\begin{align}\label{eq:indifference}
\max_{\cdf\in\Delta([0,\prior))}\max_{\continuation}\int_{[0,\prior)}\left[\frac{\prior-\belief_1}{1-\belief_1}\vh+\frac{1-\prior}{1-\belief_1}\delta\left(\payoffs(\belief_1)+\left(\frac{\belief_1}{1-\belief_1}-\frac{\prior}{1-\prior}\right)\payoffh(\belief_1)\right)\right]\cdf(d\belief_1).
\end{align}
\normalsize
\end{corollary}
The proof is in \autoref{appendix:new-b1} and is a consequence of the constraint that $\tau_0$ is Bayes' plausible for \prior. \autoref{corollary:indifference} implies that if posteriors $\belief_1$ and $\belief_1^\prime$ are on the support of $\tau_0$, then the seller is indifferent between splitting \prior\ between $\belief_1$ and $1$ and splitting \prior\ between $\belief_1^\prime$ and $1$. Indeed, given the preceding discussion, the term in the square brackets inside the integral in \autoref{eq:indifference} represents the payoff from splitting \prior\ between $1$ and $\belief_1<\prior$. \autoref{corollary:indifference} implies that, for a fixed choice of continuation payoffs, the solution to the problem in \autoref{eq:opt-1} is as if the seller were randomizing between splitting the prior between $1$ (and selling the good) and $\belief_1<\prior$ (and not selling the good). As a consequence, to determine the optimal $\tau_0$ it is enough to compare the payoffs of the splittings of \prior\ between $\belief_1$ and $1$ for different $\belief_1$. 

\section{Proof of \autoref{theorem:characterization}: key steps}\label{sec:proof}
\autoref{sec:proof} overviews the main steps of the proof of \autoref{theorem:characterization}. Taking as given that \Pbestar\ is a PBE assessment, \autoref{sec:uniqueness} shows that it achieves the unique equilibrium payoff for the seller and that, except for the threshold beliefs, $\{\delayn\}_{n\geq 1}$, there is a unique equilibrium outcome and hence, a unique equilibrium payoff for the high-valuation buyer as well. \autoref{sec:pbe-assessment} reviews the main steps to show the first statement of \autoref{theorem:characterization}; namely, that \Pbestar\ is a PBE assessment.
%
\subsection{Characterization of the equilibrium payoffs of \game\prior)}\label{sec:uniqueness}
We show that \postedpayoff(\prior) is both a lower bound and an upper bound on the seller's equilibrium payoff. That is, we show that the seller can never do better nor worse than if he were limited to choose prices in each period so that having access to a richer action space does not increase nor decrease the seller's payoff. The proof that \postedpayoff(\prior) is a lower bound is similar to the logic in \cite*{gul1986foundations}: if there is an equilibrium where the seller earns less than \postedpayoff(\prior), then he can always \emph{undercut} the price in the posted-prices assessment and earn close to \postedpayoff(\prior). The only difference is that here this undercutting is done using mechanisms. Instead, the proof that \postedpayoff(\prior) is an upper bound on the seller's equilibrium payoff takes advantage of the program in \ref{eq:opt}.

To illustrate the main steps of the proof that  \postedpayoff(\prior) is both an upper bound and a lower bound on the seller's equilibrium payoff, fix $\prior\in\Delayn$ for some $n\geq1$ and suppose that we have already shown that \autoref{theorem:characterization} holds for all $m<n$ and all $\belief\in[\delay_m,\delay_{m+1})$.\footnote{We verify that  \autoref{theorem:characterization} holds for $\prior\in[0,\delayone)$ in \autoref{appendix:opt}.} In what follows, we show that \autoref{theorem:characterization} holds for $\prior\in\Delayn$.

\paragraph{\postedpayoff(\prior) is an upper bound on the seller's equilibrium payoff:} To show that the value of \ref{eq:opt} corresponds to the seller's payoff in the posted-prices assessment, we argue that splitting the prior between $1$ and $\delaynminus$ dominates all other splittings. \autoref{corollary:indifference} implies that this is enough to show that $\postedpayoff(\prior)=\virtual(\prior)$. In what follows, we first argue why, conditional on inducing a belief $\belief_1<\delayn$, the seller places weight on at most \delaynminus\ (and on $\delay_{n-2}$ only if $\prior=\delayn$). We then argue why it is never optimal to induce a belief $\belief_1\in[\delayn,\prior)$.

To see why conditional on inducing a belief in $[0,\delayn)$,  the solution to the problem in \autoref{eq:indifference} places at most weight on $\{\delay_{n-2},\delaynminus\}$, 
recall that by the inductive hypothesis, for all $\belief_1<\delayn$, the seller's continuation payoff is as described in \autoref{theorem:characterization}. Furthermore, note the seller with prior belief \prior\ prefers the continuation payoff that minimizes the high-valuation buyer's rents whenever, conditional on no trade, he induces a posterior $\belief_1<\prior$.\footnote{This can be easily seen from noting that when $\belief_1<\prior$, the term pre-multiplying $\payoffh(\cdot)$ in \autoref{eq:virtual} is negative.} This means that at the threshold beliefs, the seller prefers to induce a continuation payoff of $\postedpayoffh(\cdot)$ for the high-valuation buyer, as is the case in the posted-prices assessment.

\autoref{lemma:vs-maximizers} shows that if the seller places positive probability on $[0,\delayn)$, then this can happen at most at $\{\delay_{n-2},\delaynminus\}$. Furthermore, $\delay_{n-2}$ can be in the support of $\tau_0$ only if $\prior=\delayn$. This is intuitive: Beliefs in $[0,\delayn)$ can be classified in terms of the time it takes to trade with the low valuation buyer, where if $\belief_1\in[\delay_m,\delay_{m+1})$, then trade happens with \vl\ in $m$ periods. It can never be optimal to induce a belief  $\belief_1\in(\delay_m,\delay_{m+1})$ for $m\leq n-1$: Both $\belief_1$ and $\delay_m$ imply that trade happens with \vl\ in $m$ periods; however, inducing $\delay_m$ allows the seller to trade with \vh\ with a higher probability. It follows that conditional on inducing beliefs in $[0,\delayn)$, the solution to \ref{eq:opt} induces at most beliefs in $\{\delay_m\}_{m\leq n-1}$. The indifference condition that defines the threshold beliefs implies that since $\prior\geq\delayn$, then it cannot be optimal to induce beliefs below $\delay_{n-2}$ (see \autoref{lemma:monotone-diff}).

To conclude the proof that $\postedpayoff(\prior)$ is an upper bound on \virtual(\prior), it remains to show that it is not optimal to induce posteriors in $[\delayn,\prior)$.
While \autoref{prop:new-b1} implies that at the solution to \ref{eq:opt} the seller's beliefs go down conditional on no trade, it does not say \emph{how fast} they go down. Indeed, it could be optimal to induce beliefs in $[\delayn,\prior)$ if there exist continuation equilibria where the seller at the induced beliefs somehow manages to slowly trade with \vh\ so as to maximally delay trade with \vl. As we show next, this cannot be optimal for \prior\ close to \delayn: The closer to \delayn, the smaller the probability that the seller with belief \prior\ can trade with \vh\ if conditional on no trade, his beliefs must remain above \delayn.
%
Thus, there exists \prior\ small enough for which it is better to trade with \vl\ in $n$ periods in exchange of increasing the probability of trading with \vh\ today.

Formally, let 
\begin{align}
F=\left\{\prior\in\Delayn:\text{\autoref{eq:delay-n} does not hold}\right\},
\end{align}
and let $\joker=\inf F$. Note that $\joker\in\Delayn$. Then, for all $\epsilon>0$, there exists $\prior\in[\joker,\joker+\epsilon)$ such that \autoref{eq:delay-n} does not hold for \prior. For each $\epsilon\in(0,\delaynplus-\delayn)$, let $\maxvirtual_\epsilon$ denote the supremum of \virtual\ over $\prior\in[\joker,\joker+\epsilon)$. For all $\prior\in[\joker,\joker+\epsilon)$, the following holds:\label{rc-game-ends}
\begin{align}\label{eq:upper bound-2}
\virtual(\prior)\leq\max\left\{\postedpayoff(\prior),\frac{\prior-\joker}{1-\joker}\vh+\frac{1-\prior}{1-\joker}\delta\maxvirtual_\epsilon\right\}.
\end{align}
To see why \autoref{eq:upper bound-2} holds note the following. First, if the solution to \ref{eq:opt} places positive mass below \joker,  \autoref{lemma:vs-maximizers} implies
%
that $\virtual(\prior)=\postedpayoff(\prior)$, since the seller places weight on \delaynminus\ (or $\delay_{n-2}$ if $\prior=\delayn$).
The equality then follows because \autoref{corollary:indifference} implies that if $\belief_1$ is in the support of $\tau_0$, then the seller's payoff coincides with the payoff he would obtain from splitting $\prior$ between $1$ and $\belief_1$. The payoff \postedpayoff(\prior) corresponds to splitting \prior\ between $1$ and $\delaynminus$. Second, if the solution to \ref{eq:opt} places weight on $[\joker,\prior)$, then the second term on the RHS of \autoref{eq:upper bound-2} is an upper bound to \virtual(\prior). After all, (i) $(\prior-\joker)/(1-\joker)$ is the largest weight that can be assigned to \vh\ while still remaining on $[\joker,\prior)$ and (ii) the remaining weight corresponds to some $\belief_1\in[\joker,\prior)$ with payoff,
\begin{align*}
\left(\payoffs(\belief_1)+(1-\belief_1)\left(\frac{\belief_1}{1-\belief_1}-\frac{\prior}{1-\prior}\right)\payoffh(\belief_1)\right)\leq\payoffs(\belief_1)\leq\maxvirtual_\epsilon,
\end{align*}
where the first inequality follows from $\belief_1<\prior$ and the second from the definition of $\maxvirtual_\epsilon$ together with $\belief_1\in[\joker,\prior)\subset[\joker,\joker+\epsilon)$. Taking the supremum over $\prior\in[\joker,\joker+\epsilon)$ on both sides of \autoref{eq:upper bound-2}, we obtain
\begin{align}\label{eq:vs-eps-upper-bound}
    \maxvirtual_\epsilon\leq\max\left\{\maxpayoff_{S\epsilon},\frac{\epsilon}{1-\joker}(\vh-\delta\maxvirtual_\epsilon)+\delta\maxvirtual_\epsilon\right\},
\end{align}
where $\maxpayoff_{S\epsilon}$ is the supremum of \postedpayoff(\prior) over $\prior\in[\joker,\joker+\epsilon)$, and the expression in the second term follows from noting that  $\prior-\joker<\epsilon$ and $\vh>\delta\maxvirtual_{\epsilon}$. 

We claim that there exists $\overline{\epsilon}>0$ such that for all $\epsilon\in(0,\overline{\epsilon})$, 
\begin{align*}
\frac{\epsilon}{1-\joker}(\vh-\delta\maxvirtual_\epsilon)+\delta\maxvirtual_\epsilon\leq\maxpayoff_{S\epsilon}.
\end{align*}
Toward a contradiction, suppose not. Then, for all $\epsilon\in(0,\delaynplus-\delayn)$, there exists $f(\epsilon)\in(0,\epsilon)$ such that 
\begin{align}\label{eq:failure}
\maxpayoff_{Sf(\epsilon)}<\frac{f(\epsilon)}{1-\joker}(\vh-\delta\maxvirtual_{f(\epsilon)})+\delta\maxvirtual_{f(\epsilon)}.
\end{align}
Since $f(\epsilon)<\epsilon$, then $f(\epsilon)\rightarrow0$ as $\epsilon\rightarrow0$. \autoref{eq:vs-eps-upper-bound} together with \autoref{eq:failure} imply that:
\begin{align*}
    \maxvirtual_{f(\epsilon)}\leq\frac{f(\epsilon)}{1-\joker}(\vh-\delta\maxvirtual_{f(\epsilon)})+\delta\maxvirtual_{f(\epsilon)},
\end{align*}
so that as $\epsilon\rightarrow0$, 
\begin{align*}
    \lim_{\epsilon\rightarrow0}\maxvirtual_{f(\epsilon)}\leq \delta\lim_{\epsilon\rightarrow0}\maxvirtual_{f(\epsilon)},
\end{align*}
a contradiction, since for all $\epsilon$,  $\maxvirtual_{\epsilon}\geq\vl>0$.\footnote{Note that the limit $    \lim_{\epsilon\rightarrow0}\maxvirtual_{f(\epsilon)}$ exists up to a convergent subnet because $\maxvirtual_{f(\epsilon)}$ is bounded. } It follows that there exists $\overline{\epsilon}>0$ such that $\forall\epsilon\in(0,\overline{\epsilon})$, 
   $ \maxvirtual_\epsilon=\maxpayoff_{S\epsilon}$.

We now claim that $\virtual(\prior)=\postedpayoff(\prior)$ for all $\prior\in[\joker,\joker+\overline{\epsilon})$. Toward a contradiction, suppose that there exists $\prior\in[\joker,\joker+\overline{\epsilon})$ such that 
$\virtual(\prior)>\postedpayoff(\prior)$. By continuity of \postedpayoff\ on $\Delayn$ (see \autoref{eq:eqbm-payoffs}), there exists $\gamma>0$ such 
that letting $\epsilon=\prior+\gamma-\joker$, we have
\begin{align*}
    \postedpayoff(\prior+\gamma)=\maxpayoff_{S\epsilon}<\virtual(\prior)\leq\maxvirtual_{\epsilon},
\end{align*}
where the first equality follows from \postedpayoff\ being increasing on $[\delayn,\delaynplus)$. This is a contradiction. 

Thus, for all $\prior\in[\joker,\joker+\overline{\epsilon})$ we have that \postedpayoff(\prior) is an upper bound on the seller's payoff.\footnote{The argument above is reminiscent to that in \cite{fudenberg1991game}'s treatment of the equilibrium in the posted-prices game with binary values.} 
%
By definition of \joker, then there exists $\prior\in[\joker,\joker+\overline{\epsilon})$ such that either (i) $\postedpayoff(\prior)$ is not the seller's unique equilibrium payoff, or (ii) there exists $\payoffb\in\eqbmset(\prior)$ such that $\payoffb_S=\postedpayoff(\prior)$, but $\payoffb_H\neq\postedpayoffh(\prior)$. Otherwise, we would have that there exists $\epsilon(=\overline{\epsilon})$ such that for all $\prior\in[\joker,\joker+\epsilon)$, \autoref{eq:delay-n} holds, which contradicts the definition of \joker.

In what follows, we first show that \postedpayoff(\prior) is a lower bound on the seller's equilibrium payoff, and hence the seller's unique equilibrium payoff. We then show that there is a unique equilibrium payoff for the high-valuation buyer, except if $\prior=\delayn$.

\paragraph{\postedpayoff(\prior) is a lower bound on the seller's equilibrium payoff (\autoref{prop:lower bound}):} Suppose there is an equilibrium of \game\prior) in which the seller earns a payoff $\minseller<\postedpayoff(\prior)$. Note that in the posted-prices equilibrium, the seller's mechanism induces posterior beliefs below \delayn, where by assumption the unique continuation payoff for the seller coincides with that of the posted-prices equilibrium. Then, if the seller were to deviate to the mechanism he offers in the posted-prices equilibrium \emph{and} we could guarantee that the buyer participates and tells the truth, the seller's continuation payoff will coincide with those in the posted-prices equilibrium. Thus, it suffices to show that the seller can guarantee that the buyer participates and tells the truth when he attempts to deviate to such a mechanism at the beginning of the game.

Key to showing that the seller can ensure such a deviation is the observation that whenever the low-valuation buyer has strict incentives to participate, equilibrium considerations imply that the high-valuation buyer also participates with probability $1$. If not, Bayes' rule would imply that the seller assigns probability $1$ to the high-valuation buyer after non-participation and thus, the high-valuation buyer earns a continuation payoff of $0$ (\autoref{lemma:vh-0}). Instead, since the low-valuation buyer strictly prefers to participate, it follows that the high-valuation buyer can ensure a strictly positive payoff from participation. Thus, the high-valuation buyer participates with probability $1$ whenever the low-valuation buyer strictly prefers to participate.

Using the above observation, we construct a mechanism that is ``close" to the one the seller offers under \Pbestar, but modified so that (i) the low-valuation buyer gets paid for participating, (ii) the high-valuation buyer gets paid for telling the truth, and (iii) instead of generating a belief of $\delaynminus$, the seller generates a belief slightly above \delaynminus. The property in (iii) ensures that the seller with belief \prior\ cannot be ``punished" by the seller with belief \delaynminus, who could choose to trade with \vl\ in $n-2$ periods.

The proof so far implies that for all $\prior\in[\joker,\joker+\overline{\epsilon})$, there is a unique equilibrium payoff for the seller and it coincides with that of the posted-prices equilibrium. It remains to show that the high-valuation buyer's payoff coincides with $\postedpayoffh(\prior)$, unless \prior=\delayn.
\paragraph{High-valuation buyer's payoff:} Suppose there exists $\payoffb\in\eqbmset(\prior)$ such that $\payoffb_S=\postedpayoff(\prior)$, but $\payoffb_H\neq\postedpayoffh(\prior)$. This equilibrium payoff, \payoffb, is associated to a mechanism in period $0$, \policynullb, and continuation payoffs, \continuationb, for the buyer and the seller. We argue that the virtual surplus associated to (\policynullb,\continuationb) must equal \postedpayoff(\prior). To see this, note that
\begin{align*}
\virtual(\policynullb,\continuationb,\prior)\geq\postedpayoff(\prior).
\end{align*}
Otherwise, we have that
\begin{align*}
\payoffb_S\leq\virtual(\policynullb,\continuationb,\prior)<\postedpayoff(\prior),
\end{align*}
where the first inequality follows from \autoref{lemma:vs-upper bound} and the second inequality is by assumption. Since $\payoffb_S=\postedpayoff(\prior)$, we have a contradiction.

Furthermore, it cannot be the case that 
\[\virtual(\policynullb,\continuationb,\prior)>\postedpayoff(\prior),\]
since \policynullb\ together with the continuation payoffs \continuationb\ are feasible choices in \ref{eq:opt}. It follows that
\[\virtual(\policynullb,\continuationb,\prior)=\postedpayoff(\prior).\]
Thus, 
\[(\policynullb,\continuationb)\in\arg\max_{\policynull,\continuation}\virtual(\policynull,\continuation,\prior).\]

However, the proof that \postedpayoff(\prior) is the value of \ref{eq:opt} implies that the solution to \ref{eq:opt} is unique, unless $\prior=\delayn$. It then follows that $\prior=\delayn$, so that there is a continuum of solutions to \ref{eq:opt}, with $\vh$'s payoff ranging from \postedpayoffh(\prior) to $\postedpayoffh(\prior)/\delta$. This completes the proof.


\subsection{ \Pbestar\ is an equilibrium assessment}\label{sec:pbe-assessment}

The analysis so far has relied on the observation that $(0,\postedpayoffh(\prior),\postedpayoff(\prior))$ is an equilibrium payoff. The rest of the proof of \autoref{theorem:characterization} shows that \Pbestar\ is an equilibrium assessment (see \autoref{appendix:equilibrium}). To do this, we first complete the equilibrium assessment by specifying the seller's and the buyer's strategy after every history (see Section \ref{appendix:complete}-\ref{appendix:sellerfull}). We then show that given beliefs and  continuation payoffs, neither the buyer nor the seller have a one shot deviation (\autoref{appendix:sellerrationality}). The results in \cite{athey2008collusion} imply that this is enough to conclude that \Pbestar\ is an equilibrium assessment. 
%
%
%

\paragraph{Seller's strategy:} Except for the cutoff beliefs $\{\delay_n\}_{n\geq1}$, we specify that the seller plays the mechanism described in the posted-prices assessment on and off the path of play. 
Instead, the seller's strategy off the path of play when his beliefs are in $\{\delayn\}_{n\geq1}$ needs to be determined \emph{jointly} with the buyer's strategy, to which we turn next.

\paragraph{Buyer's strategy (\autoref{appendix:complete}):} To complete the buyer's strategy, we first classify mechanisms according to whether they satisfy the participation and truthtelling constraints for the buyer given the continuation payoffs under \Pbestar. For mechanisms that satisfy these constraints, we specify that the buyer indeed participates and truthfully reports her value to the mechanism.

To specify the buyer's strategy for mechanisms that fail to satisfy either constraint, one needs to determine simultaneously the buyer's best response and 
the seller’s beliefs conditional on observing either the buyer reject the mechanism, or the buyer accept the mechanism and the output message that results from the buyer's report.
%
 On the one hand, the buyer's continuation payoffs depend on the seller's beliefs, which are determined by the buyer's strategy. On the other hand, whether the buyer's strategy is a best response depends on the continuation payoffs. We use the results in \cite{simon1990discontinuous} to solve for this fixed-point. It is at this point where the possibility that the seller randomizes when indifferent between trading with \vl\ in $n$ or $n-1$ periods arises to ensure that the buyer's best response is well-defined.

By specifying the buyer's strategy in the way described above, we ensure that the buyer is best responding to the seller's strategy given the continuation payoffs. It remains to show that the seller has no one-shot deviations:

\paragraph{The seller has no one-shot deviations (\autoref{appendix:sellerrationality}):} To show that the seller does not have one-shot deviations, we rely once again on the optimization problem defined by \ref{eq:opt}. For concreteness, suppose we are at a history \publict\ and let $\belief_t$ denote the seller's belief at \publict.

We show in \autoref{appendix:sellerrationality} that the payoff from any deviation at history \publict\ is bounded above by the value of the program in \ref{eq:opt} evaluated at $\prior=\belief_t$ subject to the constraint that for posterior beliefs $\belief_{t+1}$ below $\belief_t$, the continuation payoffs for the buyer and the seller are given by $(\postedpayoffh(\belief_{t+1}),\postedpayoff(\belief_{t+1}))$. The results in \autoref{sec:uniqueness} imply that the value of this program coincides with $\postedpayoff(\belief_t)$. Since this is the seller's payoff under the equilibrium strategy at \publict, it follows that the seller has no one-shot deviations at \publict.

Two observations are key to show that this upper bound holds. The first is one we have already used in the solution to \ref{eq:opt}: For threshold beliefs \delayn\ below $\belief_t$, the seller with prior belief $\belief_t$ prefers the continuation payoff vector in which the high-valuation buyer receives her lowest equilibrium payoff, \postedpayoffh(\delayn). Thus, by selecting continuation payoffs in this way, we exaggerate the payoff that the seller can guarantee from a deviation. The second is that \emph{any} mechanism \mechanismt\ together with the buyer's best response to \mechanismt\ define a new mechanism, \mechanismbt, that satisfies the participation and truthtelling constraint of the buyer \emph{given} the continuation payoffs associated to \mechanismt. Thus, we can use mechanism \mechanismbt\ together with the continuation payoffs specified by the assessment when \mechanismt\ is offered to bound the revenue from \mechanismt\ by its virtual surplus. This completes the description of the main steps in the proof of \autoref{theorem:characterization}.

\section{Conclusions}\label{sec:conclusions}
This is the first paper to characterize all equilibrium outcomes in an infinite-horizon, mechanism-selection game between an uninformed designer and a privately informed agent with persistent private information under limited commitment. We do so by marrying  insights from the literatures on bargaining and mechanism design. Following the results in our previous work, \cite{doval2020mechanism}, we endow the seller with a larger class of mechanisms that enables the seller to design his posterior beliefs about the buyer's value. The combination of mechanism design and information design elements was key in obtaining a tractable characterization. The revelation principle in \cite{doval2020mechanism}  also allows us to simplify the equilibrium behavior of the buyer, so that for the most part we were able to focus on the strategic considerations that pertain the seller.


\label{relaxed-conclusions}The study of the properties of the solution to \ref{eq:opt} can provide a useful benchmark for the analysis of the designer's best equilibrium payoff in other settings with quasilinear utility. In such settings, the virtual surplus is an upper bound on the designer's equilibrium payoff. Furthermore, when there is a continuum of types, the application of the envelope theorem implies that the designer's payoff can be represented as the virtual surplus. The program in \ref{eq:opt} is exactly like the relaxed problem in standard mechanism design: If a solution to \ref{eq:opt} satisfies the ignored constraints, then there is a PBE of the mechanism-selection game that implements the solution to \ref{eq:opt}. However, as the analysis of \cite{laffont1993theory} and, more recently, by \cite{breig2020repeated}, underscores it is not immediate that the solution to the relaxed problem satisfies \emph{all} constraints and hence, can be implemented as an equilibrium outcome. Nevertheless, like the solution to the relaxed problem in standard mechanism design, the solution to \ref{eq:opt} provides a natural benchmark to study the properties of the designer's optimal mechanism. As the analysis in this paper shows, the properties of the solution to \ref{eq:opt} can be derived even without much knowledge of how continuation equilibria look like, which reenforces its applicability (recall the results in \autoref{prop:new-b1}).

Still, unresolved issues remain for settings with a continuum of types.\footnote{In settings with finitely many types the study of \ref{eq:opt}, which is an information design problem, would reveal information about the cardinality of posteriors. For instance, Carath\'eodory's theorem (\citealp{rockafellar2015convex}) suggests that with three valuations the optimality of posted prices might be overturned because a solution to \ref{eq:opt} might use three beliefs. \label{fnt:finitely-many-types}
%
} The most important is that the solution to \ref{eq:opt} involves solving an information design problem with a continuum of states for which the existing tools do not readily apply. Indeed,  the tools in \cite{gentzkow2016rothschild}, \cite{kolotilin2018optimal} and \cite{dworczak2019simple} apply to settings where the objective function depends only on the posterior mean.
 Instead, the virtual surplus may depend on the \emph{entire} shape of the distribution, and not just the posterior mean. The characterization of equilibrium outcomes of our game  with a continuum of values remains an open question left for future research.

{\singlespacing{\bibliographystyle{ecta}
\bibliography{limited_commitment_revision,limited_commitmentVS,laura-added}}}
\appendix
\section{Omitted proofs from \autoref{sec:new-opt}}\label{appendix:omitted}
\autoref{appendix:omitted} is organized as follows. \autoref{appendix:auxiliary} introduces two auxiliary lemmas regarding the equilibrium payoffs of \vl\ (\autoref{lemma:vl-0}) and \vh\ (\autoref{lemma:vh-0}). \autoref{appendix:new-b1} presents the proof of \autoref{prop:new-b1}. 
\subsection{Auxiliary results}\label{appendix:auxiliary}
\begin{lemma}\label{lemma:vh-0}
When $\prior=1$, in any PBE of $\game\prior)$, the buyer's payoff is $0$ when her valuation is \vh.
\end{lemma}

\begin{proof}
To see that \autoref{lemma:vh-0} holds, we proceed as follows. Let $u_H^\star$ denote the highest equilibrium payoff that the buyer when her valuation is \vh\ can obtain in a PBE when \prior\ equals 1. Towards a contradiction, suppose that $u_H^\star>0$. In particular, this means that $\delta u_H^\star<u_H^\star$. Let \mechanism\ denote the mechanism offered by the seller in $t=0$.
%

Consider now a mechanism \mechanismb\ that coincides with \mechanism\ except that transfers are raised by $\epsilon>0$, where $\epsilon$ is chosen so that $\payoffh^\star-\epsilon>\delta\payoffh^\star$. We argue that the buyer must accept this mechanism with probability $1$.

Suppose that in equilibrium the buyer rejected such a mechanism. Then, Bayes' rule where possible pins down the seller's beliefs after the rejection. In particular, the buyer when her valuation is \vh\ can make at most $\delta \payoffh^\star$ in the continuation. This contradicts that it is sequentially rational to reject the mechanism. The same argument also implies that it cannot be that the buyer is indifferent between accepting and rejecting the mechanism. Thus, any such offer must be accepted with probability $1$. 

This is a contradiction because the seller's revenue under \mechanismb\ is higher than that under \mechanism, so that the seller has a deviation.
%
We conclude that $u_H^\star=0$.
\end{proof}
%
\paragraph{Proof of \autoref{lemma:vl-0}.}
\begin{proof} We establish that for any $\prior\in\Posteriors$, $\payoff\in\eqbmsetprior$ implies that $\payoff_L=0$. Toward a contradiction, let \maxl\ denote
\begin{align*}
    \sup\{u_L:(\exists\prior\in\Posteriors)(\exists (u_S,u_H)):(u_S,u_H,u_L)\in\eqbmsetprior\}.
\end{align*}
By assumption $\maxl>0$. Then, there exists a belief $\prior$ and an equilibrium payoff vector $(\payoffl,\payoffh,\payoffs)\in\eqbmsetprior$ such that $\payoffl>\delta\maxl$, by definition of the supremum. Let \Pbe\ denote the assessment that achieves that payoff. Let $\epsilon>0$ be such that $\payoffl-\epsilon>\delta\maxl$. Consider the mechanism that the seller offers, $\mechanism_0$ and consider a deviation by the seller such that he raises all payments by $\epsilon$. Since $\payoffl-\epsilon>\delta\maxl$, the best response for the low-valuation buyer is to accept this mechanism. Given this, a rejection of the mechanism reveals that the buyer's valuation is \vh. Thus, by rejecting the new mechanism, a buyer with valuation \vh\ earns at most $0$ by \autoref{lemma:vh-0}. Therefore, the buyer also accepts the mechanism when her valuation is \vh. It follows that the seller has a profitable deviation, a contradiction.
\end{proof}
\subsection{Omitted proofs from \autoref{sec:opt}}\label{appendix:new-b1}
\begin{proof}[Proof of \autoref{prop:new-b1}]
To see that part \ref{itm:no-delay-above} holds, recall that $A=[\prior,1)$. Toward a contradiction, suppose that
\begin{align*}
\int_A(1-q_0(\belief_1))\tau_0(d\belief_1)>0.
\end{align*}
Note that associated to any $(0,\payoffh(\belief_1),\payoffs(\belief_1))\in\eqbmset(\belief_1)$ chosen by the policy that solves \ref{eq:opt} there is an associated mechanism $\mechanism(\belief_1)$ and continuation payoffs,  $(0,\payoffh(\belief_2),\payoffs(\belief_2))\in\eqbmset(\belief_2)$ for the buyer and the seller that are feasible in equilibrium.\footnote{Since the set of mechanisms, \mechanismsc, is convex, it is without loss of generality to consider pure strategies for the seller which do not condition on the realization of the public randomization device. However, as in the proof that \Pbestar\ is an equilibrium assessment, it is sometimes convenient to specify the seller's strategy as if the seller were mixing.\label{fnt:pure-strategies}} Furthermore, 
\begin{align*}
&\payoffs(\belief_1)=\int_{\Posteriors}\left(\transfer^{\mechanism(\belief_1)}(\belief_2)+(1-q^{\mechanism(\belief_1)}(\belief_2))\delta\payoffs(\belief_2)\right)\tau^{\mechanism(\belief_1)}(d\belief_2)\leq\\
&\int_{\Posteriors}\left[\begin{array}{l}q^{\mechanism(\belief_1)}(\belief_2)(\belief_2\vh+(1-\belief_2)\hatv(\belief_1))+\\
(1-q^{\mechanism(\belief_1)}(\belief_2))\delta \left(\payoffs(\belief_2)+(1-\belief_2)\left(\frac{\belief_2}{1-\belief_2}-\frac{\belief_1}{1-\belief_1}\right)\payoffh(\belief_2)\right)\end{array}\right]\tau^{\mechanism(\belief_1)}(d\belief_2)\\
&=\payoffs(\belief_1)+\belief_1(\payoffh(\belief_1)-\payoffhl(\belief_1)),
\end{align*}
\normalsize
where the inequality follows from \autoref{lemma:vs-upper bound} and $\payoffhl(\belief_1)$ is defined as:
\begin{align*}
\payoffhl(\belief_1)=\int_{\Posteriors}\left[\dv q^{\mechanism(\belief_1)}(\belief_2)+(1-q^{\mechanism(\belief_1)}(\belief_2))\delta \payoffh(\belief_2)\right]\frac{1-\belief_2}{1-\belief_1}\tau^{\mechanism(\belief_1)}(d\belief_2).
\end{align*}
It can be verified that the payoff from implementing $\mechanism(\belief_1)$ in $t=0$ and choosing its associated continuation payoffs for $t=1$ yields a payoff equal to:
\begin{align}\label{eq:lower bound}
&\payoffs(\belief_1)+\belief_1\left(\payoffh(\belief_1)-\payoffhl(\belief_1)\right)+(1-\belief_1)\left(\frac{\belief_1}{1-\belief_1}-\frac{\prior}{1-\prior}\right)(\payoffh(\belief_1)-(\payoffh(\belief_1)-\payoffhl(\belief_1)))\nonumber\\
&=\payoffs(\belief_1)+(1-\belief_1)\left(\frac{\belief_1}{1-\belief_1}-\frac{\prior}{1-\prior}\right)\payoffh(\belief_1)
+(1-\belief_1)\frac{\prior}{1-\prior}(\payoffh(\belief_1)-\payoffhl(\belief_1))\nonumber\\
&\geq\payoffs(\belief_1)+(1-\belief_1)\left(\frac{\belief_1}{1-\belief_1}-\frac{\prior}{1-\prior}\right)\payoffh(\belief_1),
\end{align}
\normalsize
with the inequality being strict whenever $\payoffh(\belief_1)>\payoffhl(\belief_1)$.\footnote{One interpretation of \autoref{eq:lower bound} is the following. In the equilibrium of $\game\belief_1)$ with payoff vector $(0,\payoffh(\belief_1),\payoffs(\belief_1))$, the incentive compatibility constraint for the high valuation buyer may not have been binding. When the new policy implements the virtual surplus of the mechanism associated with $\belief_1$, two things happen. First, we increase the seller's payoff by the same amount that we decrease the buyer's payoff. Second, from today's perspective there is an additional gain: by decreasing \vh's continuation payoffs, we decrease the rents today and hence increase the virtual surplus.}

Consider the mechanism \policynullb, which coincides with \policynull\ except that whenever $\belief_1\in A$ is induced, we implement $\mechanism(\belief_1)$ in period $0$ and the continuation payoffs associated with $\mechanism(\belief_1)$ in period $1$. 
Since \policynullb\ is feasible, it must be that it is less preferred to \policynull. Using \autoref{eq:lower bound}, we have that the payoff difference between the two policies is:
\begin{align*}
&\virtual(\policynullb,(\payoffs^\prime,\payoffh^\prime),\prior)-\virtual(\policynull,\continuation,\prior)=\\
&=(1-\delta)\int_A\left(\payoffs(\belief_1)+(1-\belief_1)\left(\frac{\belief_1}{1-\belief_1}-\frac{\prior}{1-\prior}\right)\payoffh(\belief_1)\right)\tau_0(d\belief_1)
\\
&
+\int_A(1-\belief_1)\frac{\prior}{1-\prior}(\payoffh(\belief_1)-\payoffhl(\belief_1))\tau_0(d\belief_1)>0,
\end{align*}
contradicting the optimality of the original policy. 

\label{new-b1-2}To see that part \ref{itm:trade-1} in \autoref{prop:new-b1} holds, assume towards a contradiction that there exists $\belief_1<1$ such that $q_0(\belief_1)>0$. If $\belief_1>0$, consider an alternative policy which splits the weight on $\belief_1$ conditional on $q_0(\belief_1)>0$ between $0$ and $1$, and sets $q_0(0)=0$ and $q_0(1)=1$. Instead, if $\belief_1=0$, set $q_0(0)=0$. This leads to a change in payoffs of
\begin{align*}
\int_{[0,1)}q_0(\belief_1)\left[(\belief_1\vh+(1-\belief_1)\delta\hatv(\prior))-(\belief_1\vh+(1-\belief_1)\hatv(\prior))\right]\tau_0(d\belief_1)>0
\end{align*}
since $\delta<1$ and $\hatv(\prior)<0$. This is a contradiction as long as $\int_{[0,1)}q_0(\belief_1)\tau_0(d\belief_1)>0$.
\end{proof}

\autoref{prop:new-b1} implies that if (\policynull,\continuation) is a maximizer of $\virtual(\cdot,\prior)$, then
\begin{align*}
&\virtual(\policynull,\continuation,\prior)=\\
&\tau_0(\{1\})\vh+\int_{[0,\prior)}\delta\left[\frac{\prior-\belief_1}{1-\belief_1}\vh+\frac{1-\prior}{1-\belief_1}\delta\left(\payoffs(\belief_1)+\left(\frac{\belief_1}{1-\belief_1}-\frac{\prior}{1-\prior}\right)\payoffh(\belief_1)\right)\right]\tau_0(d\belief_1)\nonumber\\
&=\int_{[0,\prior)}\left[\frac{\prior-\belief_1}{1-\belief_1}\vh+\frac{1-\prior}{1-\belief_1}\delta\left(\payoffs(\belief_1)+\left(\frac{\belief_1}{1-\belief_1}-\frac{\prior}{1-\prior}\right)\payoffh(\belief_1)\right)\right]\frac{1-\belief_1}{1-\prior}\tau_0(d\belief_1)\\
&=\int_{[0,\prior)}\left[\frac{\prior-\belief_1}{1-\belief_1}\vh+\frac{1-\prior}{1-\belief_1}\delta\left(\payoffs(\belief_1)+\left(\frac{\belief_1}{1-\belief_1}-\frac{\prior}{1-\prior}\right)\payoffh(\belief_1)\right)\right]\cdf(d\belief_1)\nonumber
\end{align*}
where the second equality follows from the constraint that $\tau_0$ is Bayes' plausible for \prior\ and the third equality from noting that $(1-\belief_1)/(1-\prior)\tau_0$ defines a measure $\cdf\in\Delta([0,\prior])$. We now prove \autoref{corollary:indifference}:
\begin{proof}[Proof of \autoref{corollary:indifference}]
Any distribution over posteriors $\tau\in\Delta\Posteriors$ with support on $[0,\prior]\cup\{1\}$ induces $\cdf\in\Delta([0,\prior])$ via 
\begin{align*}
\cdf(\measurablem)=\int_{\measurablem}\frac{1-\belief_1}{1-\prior}\tau(d\belief_1).
\end{align*}
Furthermore, any distribution  $\cdf\in\Delta([0,\prior])$ induces a Bayes' plausible distribution over posteriors $\tau_0$ with support on $[0,\prior]\cup\{1\}$, where
\begin{align*}
\tau_0(\measurablem)&=\int_{\measurablem}\frac{1-\prior}{1-\belief_1}\cdf(d\belief_1),\measurablem\subset[0,\prior]\\
\tau_0(\{1\})&=\int_0^{\prior}\frac{\prior-\belief_1}{1-\belief_1}\cdf(d\belief_1).
\end{align*}
The result follows. 
\end{proof}

\section{Omitted proofs from \autoref{sec:uniqueness}}\label{appendix:uniqueness}
\autoref{appendix:uniqueness} is organized as follows. \autoref{appendix:payoffs} constructs recursively the buyer's and seller's payoffs under \Pbestar. In particular, we verify that the sequence $\{\delayn\}$ is well-defined.  Finally,  taking as given that \Pbestar\ is a PBE assessment and therefore, that the buyer and the seller's payoff under \Pbestar\ is an equilibrium payoff (that is,$(0,\postedpayoffh(\prior),\postedpayoff(\prior))\in\eqbmsetprior$), \autoref{appendix:opt} completes the proof that \autoref{eq:delay-n} holds by showing that for all $\prior\in\Posteriors$ (i) \postedpayoff(\prior) is the seller's unique equilibrium payoff for the seller, and (ii) except for the threshold beliefs, $\{\delayn\}_{n\geq1}$, there is a unique equilibrium payoff for the buyer.

\subsection{Payoffs under \Pbestar}\label{appendix:payoffs}

\paragraph{Notation:} In what follows, we denote by $(\beta_{\belief}^*,q_{\belief}^*,\transfer_{\belief}^*)$ the mechanism used by the seller with belief \belief\ in the assessment \Pbestar. 

\paragraph{Buyer's payoffs:} While the buyer's payoff is $0$ when her valuation is \vl, the buyer's payoff when her valuation is \vh\ depends on the seller's prior. Let $\payoffh^*(\prior)$ denote the buyer's equilibrium payoff in the posted prices equilibrium. We characterize this payoff inductively.

First, fix $\prior\in[0,\delay_1)$. Then, the seller sells the good at a price of \vl, so that 
\begin{align*}
\payoffh^*(\prior)=\dv.
\end{align*}

Second, fix $n\geq1$ and let $\prior\in[\delayn,\delaynplus)$. Suppose we have already shown that $\payoffh^*(\delay_m)=\delta^m\dv$ for $m\leq n-1$. Then, the buyer's payoff when her valuation is \vh\ under the equilibrium strategy is the following:
\begin{align*}
\beta_{\prior}^*(1|\vh)(\vh-(\vh-\delta^n\dv))+(1-\beta_{\prior}^*(1|\vh))\delta\payoffh^*(\delaynminus)=\delta\payoffh^*(\delaynminus)=\delta^n\dv.
\end{align*}
Note that the buyer is indifferent between reporting \vl\ and reporting \vh\ along the path of play. This is immediate for $\prior\in[0,\delay_1)$. When $\prior\in[\delayn,\delaynplus)$, the buyer's payoff from reporting \vl\ at the beginning of the game and then following her equilibrium strategy is $\delta\payoffh^*(\delaynminus)\equiv\delta^n\dv$. Thus, along the path of play, the low-valuation buyer is indifferent between participating in the mechanism and not, whereas the high-valuation buyer is indifferent between reporting the truth and not. 

\paragraph{Seller's payoffs:} Consider a seller with belief $\prior\in\Delayn$. The mechanism used by the seller determines $\policyprior$ defined as follows. First, if $\prior\in[0,\delayone)$, we have that $(\tau_{\prior}^*(\{\prior\}),q_{\prior}^*(\prior))=(1,1)$. Second, if $\prior\in\Delayn$ for $n\geq1$
\begin{align}
\tau_{\prior}^*(1)&=\frac{\prior-\delaynminus}{1-\delaynminus}=1-\tau_{\prior}^*(\delaynminus)\\
q_{\prior}^*(1)&=1-q_{\prior}^*(\delaynminus).
\end{align}

%
%
To define the seller's payoff, we first define the cutoffs $\{\delay_n\}_{n\geq 0}$ and the seller's payoff in the equilibrium when his belief is $\delay_n$.

For $n=0$, let
\begin{align*}
\postedpayoff(\prior)=\vl=\prior\vh+(1-\prior)\hatv(\prior).
\end{align*}
Define \delayone\ as the belief $\belief\in[0,1]$ such that
\small
\begin{align}\label{eq:delay-one-indiff}
\delayone\vh+(1-\delayone)\delta\left[\postedpayoff(0)+(1-0)\times(\frac{0}{1-0}-\frac{\delayone}{1-\delayone})\payoffh^*(0)\right]=\delayone\vh+(1-\delayone)\hatv(\delayone)
\end{align}
\normalsize
and let $\postedpayoff(\delayone)$ denote the value of the expression in \autoref{eq:delay-one-indiff}.
Recursively, suppose we have defined $\postedpayoff(\delay_m)$ for $1\leq m\leq n-1$. Define $\delayn$ to be the belief $\belief\in[0,1]$ such that
\small
\begin{align}\label{eq:deltanplus}
&\frac{\delayn-\delaynminus}{1-\delaynminus}\vh+\frac{1-\delayn}{1-\delaynminus}\delta\left[\postedpayoff(\delaynminus)+(1-\delaynminus)\left(\frac{\delaynminus}{1-\delaynminus}-\frac{\delayn}{1-\delayn}\right)\payoffh^*(\delaynminus)\right]\\
&=\frac{\delayn-\delay_{n-2}}{1-\delay_{n-2}}\vh+\frac{1-\delayn}{1-\delay_{n-2}}\delta\left[\postedpayoff(\delay_{n-2})+(1-\delay_{n-2})\left(\frac{\delay_{n-2}}{1-\delay_{n-2}}-\frac{\delayn}{1-\delayn}\right)\payoffh^*(\delay_{n-2})\right],\nonumber
\end{align}
\normalsize
and define $\postedpayoff(\delayn)$ to be the LHS of \autoref{eq:deltanplus}.

\autoref{lemma:monotone-diff} implies that the cutoffs determined by \autoref{eq:deltanplus}  are well-defined and satisfy $0<\delayone<\dots<\delayn<\delaynplus<\dots$:

\begin{lemma}\label{lemma:monotone-diff}
The sequence of thresholds defined by \autoref{eq:deltanplus} is increasing.
\end{lemma}
\begin{proof}[Proof of \autoref{lemma:monotone-diff}]
Recall that we are defining $\delay_0=0$ and $\delayone=\nicefrac{\vl}{\vh}$. Suppose we have shown that $\delaynminus>\delay_{n-2}$. We show that $\delayn>\delaynminus$. 

To do so, fix $\belief\geq\delaynminus$. We claim that the difference:
\small
\begin{align}\label{eq:delta-n}
&\Delta_n(\belief;\delaynminus,\delay_{n-2})=\frac{\belief-\delaynminus}{1-\delaynminus}\vh+\frac{1-\belief}{1-\delaynminus}\delta\left[\postedpayoff(\delaynminus)+(1-\delaynminus)\left(\frac{\delaynminus}{1-\delaynminus}-\frac{\belief}{1-\belief}\right)\payoffh^*(\delaynminus)\right]\nonumber\\
&-\left(\frac{\belief-\delay_{n-2}}{1-\delay_{n-2}}\vh+\frac{1-\belief}{1-\delay_{n-2}}\delta\left[\postedpayoff(\delay_{n-2})+(1-\delay_{n-2})\left(\frac{\delay_{n-2}}{1-\delay_{n-2}}-\frac{\belief}{1-\belief}\right)\payoffh^*(\delay_{n-2})\right]\right)
\end{align}
\normalsize
is increasing in $\belief$. Note that $\Delta_n$ is differentiable in $\belief$, and
\small
\begin{align*}
&\frac{\partial}{\partial\belief}\Delta_n(\belief;\delaynminus,\delay_{n-2})
=\frac{\vh(\delaynminus-\delay_{n-2})}{(1-\delaynminus)(1-\delay_{n-2})}-\delta \vh\frac{\delaynminus-\delay_{n-2}}{(1-\delaynminus)(1-\delay_{n-2})}+\delta(\delta^{n-2}-\delta^{n-1})\vh>0.
\end{align*}
\normalsize
\newline\indent
The cutoff $\delayn$ is defined by $\Delta_n(\delayn;\delaynminus,\delay_{n-2})=0$. Note that $\delayn\neq\delaynminus$ if $n\geq 1$. If $\delayn=\delaynminus$, then 
\begin{align*}
0=\Delta_n(\delayn;\delaynminus,\delay_{n-2})=\delta\postedpayoff(\delaynminus)-\postedpayoff(\delaynminus)<0,
\end{align*}
since $\delta<1$. Because $\Delta_n(\cdot;\delaynminus,\delay_{n-2})$ is increasing, we conclude that $\delayn>\delaynminus$.
\end{proof}

Having established this, we can now define the seller's payoff at any history \publict\ under \Pbestar. If $\belief^*(\publict)\in \Delayn$, then
\small
\begin{align}\label{eq:eqbm-payoffs}
\Payoffs^*(\publict)&=\frac{\belief^*(\publict)-\delaynminus}{1-\delaynminus}\vh+\frac{1-\belief^*(\publict)}{1-\delaynminus}\delta\left[\postedpayoff(\delaynminus)+(1-\delaynminus)\left(\frac{\delaynminus}{1-\delaynminus}-\frac{\belief^*(\publict)}{1-\belief^*(\publict)}\right)\payoffh^*(\delaynminus)\right]\nonumber\\
&\equiv\postedpayoff(\belief^*(\publict))
\end{align}
\normalsize
In what follows, we simplify notation by denoting for any $\prior,\belief_1\in\Posteriors$:
\begin{align}\label{eq:adjusted-payoffs}
\Req(\belief_1,\prior)=\postedpayoff(\belief_1)+(1-\belief_1)\left(\frac{\belief_1}{1-\belief_1}-\frac{\prior}{1-\prior}\right)\payoffh^*(\belief_1),
\end{align}
and note that $\Req(\prior,\prior)=\postedpayoff(\prior)$. Furthermore, if $\prior\in\Delayn$, we can write $\Req(\prior,\prior)$ as
\begin{align*}
\Req(\prior,\prior)=\frac{\prior-\delaynminus}{1-\delaynminus}\vh+\frac{1-\prior}{1-\delaynminus}\delta\Req(\delaynminus,\prior).
\end{align*}

\autoref{eq:deltanplus} implies that under the specification of equilibrium play, when the seller's belief is \delayn, he is indifferent between a posted price of $\vl+(1-\delta^n)\dv$ and a price of $\vl+(1-\delta^{n-1}\dv)$. This, in turn, implies that the buyer when her payoff is \vh\ may obtain any payoff in $[\delta^n\dv,\delta^{n-1}\dv]$, if the seller were to randomize between these two posted prices. This randomization is important for the specification of the buyer and the seller's strategies off the path of play. For future reference, let $\pazocal{U}_{H}^*$ denote the following correspondence:
\begin{align}\label{eq:correspondence}
\pazocal{U}_{H}^*(\prior)=\left\{\begin{array}{cl}u_H^*(\prior)&\text{if }\prior\neq\delay_n,n\geq1\\
\left[\delta^{n}\dv,\delta^{n-1}\dv\right]&\text{if }\prior=\delay_n\text{ for some }n\geq 1\end{array}\right..
\end{align}
Note $\pazocal{U}_{H}^*$ is upper-hemicontinuous, convex-valued, and compact-valued.

\subsection{Omitted proofs from \autoref{sec:uniqueness}}\label{appendix:opt}
\autoref{appendix:opt} completes the steps to show that the set of equilibrium payoffs of \game\prior) is as described in \autoref{eq:delay-n}. In what follows, we first establish that \autoref{eq:delay-n} holds for $\prior<\delayone$. We then establish that \postedpayoff(\prior) is a lower bound on the seller's equilibrium payoff (\autoref{prop:lower bound}). Finally, we provide the omitted proofs of the statements in \autoref{sec:uniqueness} regarding the properties of the solutions to \ref{eq:opt} (\autoref{lemma:vs-maximizers}).

\paragraph{\autoref{eq:delay-n} holds for $\prior<\delayone$:} This case is simple. First we need to show that the seller's highest equilibrium payoff is \vl\ for $\prior<\delayone$ which is immediate since this is the seller's payoff in the commitment solution and there exists a PBE assessment \Pbestar\ which achieves it. Very similar arguments to those in \autoref{prop:lower bound} imply that the seller's payoff cannot be below \postedpayoff(\prior). 

\paragraph{\autoref{eq:delay-n} holds for $\prior\geq\delayone$:} In the rest of this section, we focus on the case in which $\prior\geq\delayone$. As in the main text, let $n\geq1$ denote the smallest $m$ such that there exists $\prior\in[\delay_m,\delay_{m+1})$ such that \autoref{eq:delay-n} does not hold. To complete the proof in \autoref{sec:uniqueness}, we first show that \postedpayoff(\prior) is a lower bound on the seller's equilibrium payoff (\autoref{prop:lower bound}). We then complete the steps to show that when $\prior\in\Delayn$, the solution to \ref{eq:opt} places weight at most on $\{\delay_{n-2},\delaynminus\}$ (\autoref{lemma:vs-maximizers}):

\autoref{prop:lower bound} shows that the seller can always guarantee \postedpayoff(\prior):
\begin{prop}[\postedpayoff(\prior) is a lower bound]\label{prop:lower bound}
Let $n\geq 1$ denote the minimum $m\geq 1$ such that there exists $\prior\in [\delay_m,\delay_{m+1})$ such that \autoref{eq:delay-n} fails to hold. Let $\prior\in\Delayn$ and let $\payoff\in\eqbmsetprior$. Then, $\payoffs\geq\postedpayoff(\prior)$.
\end{prop}
\begin{proof} 
Toward a contradiction, let $\underline{\payoff}\equiv(0,\underline{\payoff}_H,\minseller)\in\eqbmsetprior$ be such that $\minseller<\postedpayoff(\prior)$.
%
%
%
Fix $\epsilon,\gamma>0$ and consider the following mechanism, \mechanismzb:
 \begin{align*}
 \betazb(\delaynminus+\gamma|\vh)&=\frac{\delaynminus+\gamma}{\prior}\frac{1-\prior}{1-(\delaynminus+\gamma)}, \betazb(1|\vh)=\frac{1}{\prior}\frac{\prior-(\delaynminus+\gamma)}{1-(\delaynminus+\gamma)}\\
 \betazb(\delaynminus+\gamma|\vl)&=1\\
 \qzb(1)&=1,\tzb(1)=\vl+(1-\delta^n)\dv-\epsilon,\\\qzb(\delaynminus+\gamma)&=0,\tzb(\delaynminus+\gamma)=-\epsilon/2.
 \end{align*}
Pick $\gamma$ so that $\delaynminus+\gamma\in[\delaynminus,\delayn)$. Then, the seller's continuation payoff is $\postedpayoff(\delaynminus+\gamma)$. 

We claim that equilibrium considerations imply that the buyer has to accept \mechanismzb\ with probability one for both of her values. By \autoref{lemma:vl-0}, the low-valuation buyer obtains a payoff of $0$ by rejecting the mechanism. Instead, by participating the buyer can guarantee a payoff of $\nicefrac{\epsilon}{2}$. Thus, sequential rationality of the buyer's strategy when her value is \vl\ implies that $\pi_{\vl}(\mechanismzb)=1$. Suppose then that $\vh$ in equilibrium rejects such an offer. In this case, Bayes' rule where possible implies that the seller must assign probability $1$ to $\vh$, in which case the buyer's continuation value is $0$ (\autoref{lemma:vh-0}). Then, it cannot be a best response to reject the mechanism with positive probability since $\nicefrac{\epsilon}{2}>0$ is a lower bound on the buyer's payoff from participating in the mechanism. Thus, in the PBE assessment that sustains $\underline{\payoff}$ the buyer must accept \emph{any} such mechanism and beliefs must be specified so that participation is indeed a best response. Note that the latter is always feasible since assigning probability one to the rejection coming from \vh\ justifies $0$ continuation values.

Furthermore, as long as $\epsilon<2(1-\delta^n)\dv$, the above mechanism gives the buyer a strict incentive to tell the truth. Thus, the seller's payoff from a deviation is lower bounded by
\begin{align*}
&-\epsilon+\frac{\prior-(\delaynminus+\gamma)}{1-(\delaynminus+\gamma)}(\vl+(1-\delta^n)\dv)+\frac{1-\prior}{1-(\delaynminus+\gamma)}\delta \postedpayoff(\delaynminus+\gamma)=\\
&-\epsilon+\frac{\prior-(\delaynminus+\gamma)}{1-(\delaynminus+\gamma)}\vh+\frac{1-\prior}{1-(\delaynminus+\gamma)}\times\\
&\delta\left[\postedpayoff(\delaynminus+\gamma)+(1-(\delaynminus+\gamma))\left(\frac{\delaynminus+\gamma}{1-(\delaynminus+\gamma)}-\frac{\prior}{1-\prior}\right)\delta^{n-1}\dv\right]\\
&=-\epsilon+[\frac{\prior-(\delaynminus+\gamma)}{1-(\delaynminus+\gamma)}\vh+\frac{1-\prior}{1-(\delaynminus+\gamma)}\delta\Req(\delaynminus+\gamma,\prior)].
\end{align*}
Note that at $\gamma=0$, the above is precisely $\postedpayoff(\prior)-\epsilon$ (recall \autoref{eq:adjusted-payoffs}), so that for $\epsilon,\gamma$ small enough we have that 
\begin{align*}
-\epsilon+[\frac{\prior-(\delaynminus+\gamma)}{1-(\delaynminus+\gamma)}\vh+\frac{1-\prior}{1-(\delaynminus+\gamma)}\delta\Req(\delaynminus+\gamma,\prior)]>\minseller.
\end{align*}
 Thus, the seller has a profitable deviation. This contradicts that there is a PBE assessment where the seller earns less than $\postedpayoff(\prior)$.
 \end{proof}

To conclude the proof that \postedpayoff(\prior) is an upper bound on the seller's payoff, we prove \autoref{lemma:vs-maximizers}. As in the main text, let $\joker=\inf\{\prior\in\Delayn:\text{\autoref{eq:delay-n} does not hold}\}$. Thus, for $\belief_1\in[0,\joker)$, the seller's payoffs are given by $\postedpayoff(\belief_1)$. Furthermore, since $\prior\geq\delayn$, the seller prefers to minimize the buyer's continuation payoff at $\{\delay_m\}_{m\leq n}$. This implies that the objective function in \autoref{eq:indifference} can be written as:
\begin{align}\label{eq:opt-2}
&\int_{[0,\joker)}\left[\frac{\prior-\belief_1}{1-\belief_1}\vh+\frac{1-\prior}{1-\belief_1}\delta\underbrace{\left(\postedpayoff(\belief_1)+\left(\frac{\belief_1}{1-\belief_1}-\frac{\prior}{1-\prior}\right)\postedpayoffh(\belief_1)\right)}_{\Req(\belief_1,\prior)}\right]\cdf(d\belief_1)\\
&+\int_{[\joker,\prior)}\left[\frac{\prior-\belief_1}{1-\belief_1}\vh+\frac{1-\prior}{1-\belief_1}\delta\left(\payoffs(\belief_1)+\left(\frac{\belief_1}{1-\belief_1}-\frac{\prior}{1-\prior}\right)\payoffh(\belief_1)\right)\right]\cdf(d\belief_1)\nonumber
\end{align}
\begin{lemma}\label{lemma:vs-maximizers}
Suppose \cdf\ maximizes the expression in \autoref{eq:opt-2} and is such that $G([0,\joker))>0$. Then, $G$ places positive probability on at most $\{\delay_{n-2},\delaynminus\}$.
\end{lemma}

\begin{proof}
It is immediate to see that no $\belief_1\in\cup_{m=0}^{n-1}(\delay_m,\delay_{m+1})$ can be on the support of \cdf: if $\belief_1\in[\delay_m,\delay_{m+1})$ for $m\leq n-1$, this is dominated by choosing $\delay_m$:\footnote{While the algebra is tedious, the intuition is immediate: $\belief_1$ in $(\delay_m,\delay_{m+1})$ generate the same probability of trading with \vl\ as $\delay_m$, but a lower probability of trading with \vh.} 
\begin{align*}
&\frac{\prior-\delay_m}{1-\delay_m}\vh+\frac{1-\prior}{1-\delay_m}\delta\Req(\delay_m,\prior)-\left(\frac{\prior-\belief_1}{1-\belief_1}\vh+\frac{1-\prior}{1-\belief_1}\delta\Req(\belief_1,\prior)\right)\\
&=\vh\left[\frac{\prior-\delay_m}{1-\delay_m}+\delta\frac{1-\prior}{1-\delay_m}\frac{\delay_m-\delay_{m-1}}{1-\delay_{m-1}}-\frac{\prior-\belief_1}{1-\belief_1}-\delta\frac{1-\prior}{1-\belief_1}\frac{\belief_1-\delay_{m-1}}{1-\delay_{m-1}}\right]>0
\end{align*}
A similar argument implies that $\belief_1\in[\delayn,\joker)$ is dominated by choosing $\delaynminus$. Thus, if it is optimal to set $\cdf([0,\joker))>0$, we can reduce the problem of finding the optimal such \cdf\ to
\begin{align}\label{eq:new-opt}
\max_{\cdf\in\Delta(\{\delay_0,...,\delay_{n-1}\})}\sum_{m=0}^{n-1}\left[\frac{\prior-\delay_m}{1-\delay_m}\vh+\frac{1-\prior}{1-\delay_m}\delta\Req(\delay_m,\prior)\right]\cdf(\delay_m)
\end{align}
However, for $m\leq n-2$, \autoref{lemma:monotone-diff} implies that
\small
\begin{align*}
&\Delta_{m+1}(\prior;\delay_m,\delay_{m-1})\\
&=\left[\frac{\prior-\delay_m}{1-\delay_m}\vh+\frac{1-\prior}{1-\delay_m}\delta\Req(\delay_m,\prior)\right]-\left[\frac{\delay_{m-1}-\prior}{1-\delay_{m-1}}\vh+\frac{1-\prior}{1-\delay_{m-1}}\delta\Req(\delay_{m-1},\prior)\right]>0,
\end{align*}
\normalsize
since $\prior>\delaynminus$. Thus, any solution to the problem in \autoref{eq:new-opt} satisfies that there exists some $\alpha\in[0,1]$ such that the value of this problem is given by:
\small
\begin{align*}
&=\alpha\left[\frac{\prior-\delaynminus}{1-\delaynminus}\vh+\frac{1-\prior}{1-\delaynminus}\delta\Req(\delaynminus,\prior)\right]+(1-\alpha)\left[\frac{\prior-\delay_{n-2}}{1-\delay_{n-2}}\vh+\frac{1-\prior}{1-\delay_{n-2}}\delta\Req(\delay_{n-2},\prior)\right]\\
&=\alpha\Delta_n(\prior;\delaynminus,\delay_{n-2})+\left[\frac{\prior-\delay_{n-2}}{1-\delay_{n-2}}\vh+\frac{1-\prior}{1-\delay_{n-2}}\delta\Req(\delay_{n-2},\prior)\right]
\end{align*}
\normalsize
so that unless $\prior=\delayn$ it is not optimal to set $\alpha<1$. Instead, for $\prior=\delayn$ any $\alpha\in[0,1]$ is a maximizer.
\end{proof}

The argument in \autoref{sec:uniqueness} implies that if $\prior\in[\delayn,\delaynplus)$, then any \cdf\ that maximizes the expression in \autoref{eq:opt-2} is such that $\cdf([\delayn,\prior))=0$. Thus, the value of \ref{eq:opt} is \postedpayoff(\prior). This completes the proof that the seller's equilibrium payoff is unique and coincides with that under \Pbestar. 
\section{\Pbestar\ is an equilibrium assessment}\label{appendix:equilibrium}
To complete the proof of \autoref{theorem:characterization}, it remains to show that \Pbestar\ is an equilibrium assessment. First, 
%
 we finalize the construction of the assessment \Pbestar, by constructing the buyer's (\autoref{appendix:complete}) and seller's strategy profile and system of beliefs (\autoref{appendix:sellerfull}) after every history. 
 \autoref{appendix:sellerrationality} shows that neither the seller nor the buyer have one-shot deviations from the equilibrium strategy, given the continuation values implied by \Pbestar. The results in \cite{athey2008collusion} imply that this is enough to conclude we have indeed constructed a PBE of $\game\prior)$.
 
 In what follows, to simplify notation we denote by $\gamma^*(\belief)$ the mechanism $(\beta_{\belief}^*,q_{\belief}^*,\transfer_{\belief}^*)$ that the seller uses in the posted-prices assessment.
\subsection{Completing the buyer's strategy}\label{appendix:complete}

To complete the buyer's strategy, for a mechanism \mechanism\ let
\begin{align}\label{eq:value-participation}
U_{1v}(\mechanism)=\max_{\rho\in\Posteriors}\sum_{v^\prime\in V}\rho(v^\prime)\int_{\Posteriors\times\allocations}(vq-\transfer)\varphi^{\mechanism}(d(\belief,q,\transfer)|v^\prime),
\end{align}
denote the buyer's maximum payoff when her value is $v$ from participating in mechanism \mechanism, without taking into account her continuation payoffs. 

Fix a public history \publict\ and let $\belief_t$ denote the seller's beliefs at that public history.\footnote{The belief, $\belief_t$, is an equilibrium object, but we suppress this from the notation to keep things simple.} To complete the buyer's strategy, we classify mechanisms, $\mechanismc$, in four categories:\footnote{\cite{gerardi2020dynamic} use a similar trick to complete the worker's strategy in their paper.}
\begin{enumerate}\setcounter{enumi}{-1}
\item\label{itm:zero} Mechanisms that given the continuation values, $\payoffh^*(\cdot)$, satisfy participation and truthtelling.
Denote the set of these mechanisms $\mechanismsc^0$.
\item Mechanisms \mechanism\ not in $\mechanismsc^0$ such that $U_{1,\vh}(\mechanism)<0$. Let
%
$\mechanismsc^1$ denote the set of these mechanisms.
\item Mechanisms not in $\mechanismsc^0$ such that $U_{1\vh}(\mechanism)\geq0>U_{1\vl}(\mechanism)$.  Let $\mechanismsc^2$ denote the set of these mechanisms.
\item Mechanisms not in $\mechanismsc^0$ such that $U_{1v}(\mechanism)\geq0$ for $v\in\{\vl,\vh\}$. Let $\mechanismsc^3$ denote the set of these mechanisms.
\end{enumerate}
If $\mechanismc\in\mechanismsc^1$, specify that the buyer rejects the mechanism for both of her values. Hence, under this strategy, the seller does not update his beliefs after observing a rejection. If, however, the buyer accepts, the seller believes $v=\vh$. Note that, in this case, continuation payoffs for the buyer are 0 from then on, regardless of her value. For each type $v$, let $r_v^*(\mechanismc,1)$ denote a maximizer of \autoref{eq:value-participation}.\footnote{Even if the buyer does not participate on the equilibrium path, we still need to guarantee the reporting strategy is sequentially rational.}

If $\mechanismc\in\mechanismsc^2$, specify that the buyer rejects \mechanism\ when her value is $\vl$. Hence, without loss of generality, we can specify that if the seller observes that the buyer accepts the mechanism, the buyer's value is $\vh$. For $\vh$, let $r_{\vh}^*(\mechanismc,1)$ denote a maximizer of \autoref{eq:value-participation} for $v=\vh$.
Note we are using that the buyer's continuation payoffs are $0$ conditional on her accepting the mechanism, so that $U_{1\vh}(\mechanism)$ is indeed the utility of the buyer when her value is \vh\ and the seller offers mechanism \mechanism\ in the game. Then, $\vh$'s payoff from participating in the mechanism $\mechanismc$ is given by $U_{1\vh}(\mechanism)$
whereas the payoff from rejecting is
$U_{0,\vh}(\pi_{\vh},f)=\delta f(\nu_2(\belief_t,\pi_{\vh})),$
where
\begin{align*}
\nu_2(\belief_t,\pi_{\vh})=\frac{\belief_t(1-\pi_{\vh})}{\belief_t(1-\pi_{\vh})+1-\belief_t}.
\end{align*}
is the seller's belief that the buyer's value is $\vh$ when observing a rejection, according to Bayes' rule, and $f(\nu_2(\belief_t,\pi_{\vh}))$ is a measurable selection from $\pazocal{U}_{H}^*(\nu_2(\belief_t,\pi_{\vh}))$. Note the payoff from rejecting is specified under the assumption that in the continuation, the equilibrium path coincides with that of the posted prices equilibrium when beliefs are $\nu_2(\belief_t,\pi_{\vh})$. We use, however, a selection from $\pazocal{U}_{H}^*$ to ensure that if needed, the seller randomizes between the posted prices when indifferent to help make the buyer's continuation problem well-behaved. 
\newline\indent
Now, $(\pi_{\vh}^*(\mechanism),f_2^*(\mechanism))$ are chosen so that 
\begin{align*}
\pi_{\vh}^*\in \arg\max_{p\in[0,1]}(1-p)U_{0,\vh}(\pi_{\vh}^*,f_2^*(\mechanism))+p U_{1,\vh}(\mechanismc)
\end{align*}
The main result in \cite{simon1990discontinuous} implies a solution to the above problem exists, given the properties of $\pazocal{U}_{H}^*$ and the linearity in $p$ of the objective. Let $\pi_{\vh}^*(\mechanismc)$ denote this fixed point. Now, if $\pi_{\vh}^*(\mechanism)<1$ and $\nu_2(\belief_t,\pi_{\vh}^*(\mechanism))=\delay_i$ for some $i\geq1$,  then there exists a weight $\phi_2(\delay_i,\mechanism)\in[0,1]$ that solves the following:
\begin{align}\label{eq:weight2}
f_2^*(\mechanism,\delay_i)=(1-\phi)\delta^{i-1}\dv+\phi\delta^i\dv.
\end{align} 
This weight captures the probability with which the seller, when his belief is $\delay_i$, mixes between $\gamma^*(\delay_i)$ and a mechanism that splits the prior $\delay_i$ between $1$ and $\delay_{i-2}$, which we denote by $\gamma^{**}(\delay_i)$.\footnote{Formally, if $i=1$, $\gamma^{**}(\delay_i)$ is the mechanism that sells the good with probability $1$ at a price of \vl, that is,  $\betam(\delay_i|\vl)=\betam(\delay_i|\vh)=1$ and $(\qm(\delay_i),\transferm(\delay_i)=(1,\vl)$. For $i\geq 2$, $\gamma^{**}(\delay_i)$ is the mechanism such that $\betam(\delay_{i-2}|\vl)=1$, $\betam(\delay_{i-2}|\vh)=(\nicefrac{\delay_{i-2}}{\delay_i})(\nicefrac{1-\delay_i}{1-\delay_{i-2}})=1-\betam(1|\vh)$ and sets $(\qm(\delay_{i-2}),\tm(\delay_{i-2}))=(0,0)$ and $(\qm(1),\tm(1))=(1,\vl+(1-\delta^{i-1})\dv)$.} For a mechanism in $\mechanismsc^2$, let $r_{\vl}^*(\mechanismc,1)$ denote a solution to \autoref{eq:value-participation} for $v=\vl$.

Finally, if $\mechanismc\in\mechanismsc^3$, specify that the buyer participates for both her values. If the seller observes that the buyer rejects the mechanism, he assigns probability $1$ to the buyer's valuation being $\vh$. Thus, upon rejection, the buyer's continuation payoff is $0$ independently of her value. Note that mechanisms in $\mechanismsc^3$ satisfy the participation constraint given the continuation values of the posted-prices assessment. Now, let $\ml$ satisfy\footnote{We cannot ensure that truthtelling will hold for mechanisms in $\mechanismsc^3$, which is why we need the extra piece of notation.} 
\begin{align*}
\int_{\Posteriors\times\allocations}(\vl q-\transfer)\varphim(d(\belief_{t+1},q,\transfer)|\ml)\geq\int_{\Posteriors\times\allocations}(\vl q-\transfer)\varphim(d(\belief_{t+1},q,\transfer)|v),
\end{align*}
for all $v\in V$. Set $r_{\vl}^*(\mechanism,1)(\ml)=1$. Let $\{\mh\}=V\setminus\{\ml\}$ and define
$\Posteriors^H\equiv\text{supp }\varphim(\cdot\times\allocations|\mh)$, $\Posteriors^L\equiv\text{supp }\varphim(\cdot\times\allocations|\ml).$
Let $r\in[0,1]$ denote the weight the buyer assigns to $\ml$ when her valuation is $\vh$. Given this notation, let $\nu_3(\belief_t,\belief_{t+1},r)$ denote the seller's belief that the buyer's valuation is \vh\ when he observes output message $\belief_{t+1}$. We assume that if $\belief_{t+1}$ is not consistent with $m_L^*$, i.e., $\belief_{t+1}\in\Posteriors^H\setminus\Posteriors^L$, then $\nu_3(\belief_t,\belief_{t+1},r)=1$.
%
This specification of beliefs does not conflict with Bayes' rule where possible: either $\mh$ has positive probability in the optimal reporting strategy of the buyer when her valuation is $\vh$, in which case, $\nu_3$ would be consistent with Bayes' rule, or it does not, in which case, Bayes' rule where possible places no restrictions on $\nu_3(\belief_t,\belief_{t+1},\cdot)$ for $\belief_{t+1}\in\Posteriors^H\setminus\Posteriors^L$.
\newline\indent
Given $\nu_3$, the buyer when her valuation is $\vh$ obtains a payoff of
\begin{align*}
\hat{U}_{1,\vh}(r,m,f)=\int_{\Posteriors\times\allocations}(\vh q-\transfer+\delta(1-q)f(\nu_3(\belief_t,\belief_{t+1},r)))\varphim(d(\belief_{t+1},q,\transfer)|m),
\end{align*}
when she reports $m\in V$, where $f$ is a selection from $\pazocal{U}_{H}^*$. We want to find $(r_{\vh}^*,f_3^*(\mechanism))$ so that
\begin{align}
r_{\vh}^*\in\arg\max_{r\in[0,1]}r\hat{U}_{1,\vh}(r_{\vh}^*,\ml,f_3^*(\mechanism))+(1-r)\hat{U}_{1,\vh}(r_{\vh}^*,\mh,f_3^*(\mechanism)).
\end{align}
The main theorem in \cite{simon1990discontinuous} implies the existence of such an $(r_{\vh}^*,f_3^*(\mechanism))$. Set $r_{\vh}^*(\mechanism,1)(\ml)=r_{\vh}^*$.
 As we did before, whenever $\nu_3(\belief_t,\belief_{t+1},r_{\vh}^*)=\delay_i$ for $i\geq1$, we can define $\phi_3(\mechanism,\delay_i)$ as the weight on $\gamma^*(\delay_i)$ implied by $f_3^*(\mechanism,\delay_i)$.
\newline\indent
Summing up, the buyer's strategy at any history $\hagt$ is given by
\begin{align}\label{eq:participationeqbm}
\pi_{tv}^*(\hagt,\mechanismc)=\left\{\begin{array}{ll}1&\text{if }\mechanismc\in\mechanismsc^0\cup\mechanismsc^3\\\pi_{\vh}^*(\mechanismc)&\text{if }\mechanismc\in\mechanismsc^2 \text{ and } v=\vh\\0&\text{otherwise}\end{array}\right.
\end{align}
and 
\begin{align}\label{eq:reportingeqbm}
r_{tv}^*(\hagt,\mechanismc,1)&=\left\{\begin{array}{ll}\delta_v&\text{ if }\mechanismc\in\mechanismsc^0\\r_v^*(\mechanismc,1)&\text{otherwise}\end{array}\right..
\end{align}
\subsection{Full specification of the PBE assessment}\label{appendix:sellerfull}
To complete the PBE assessment for $\game\prior)$, we now specify the seller's strategy and the belief system. We introduce two pieces of notation: the first one allows us to keep track of the last payoff-relevant event; the second one allows us to keep track of how the seller's beliefs evolve given the buyer's strategy (\autoref{eq:mapbeliefs}).
\newline\indent
From the beginning of period $t$ of the game until the end of that period, the following things are determined: (i) the seller's choice of mechanism, $\mechanismc$; (ii) the buyer's participation decision, $\participate\in\{0,1\}$; and (iii) if the buyer participates in the mechanism, the allocation and the output message. 
Let \znpt\ denote the tuple $(\mechanismt,0,\emptyset,(0,0))$ and $z_{(\posterior,(q,\transfer))}(\mechanismt)$ denote the tuple $(\mechanismt,1,\posterior,(q,\transfer))$.
%
%
%
Of course, if $q=1$, the game ends. Note that any public history $h^\dateindex$ can be written as $(h^{\dateindex-1},z)$ for some $z$ as defined above. Given $h^\dateindex$, let $z(h^\dateindex)$ denote the corresponding outcome $z$.
Given any prior $\prior$, let $T(\prior,z)$ denote the map that associates to each prior belief \prior\ and each outcome $z$, a seller's belief that the buyer's valuation is \vh. Formally, 
\begin{align}\label{eq:mapbeliefs}
T(\prior,z)&=\left\{\begin{array}{ll}\posterior&\text{if }z=z_{(\posterior,\cdot)}(\mechanismc)\text{ and }\mechanismc\in\mechanismsc^0\\1&\begin{array}{c}\text{if }z=z_\emptyset(\mechanismc)\text{ and }\mechanismc\in\mechanismsc^0\cup\mechanismsc^3\\\text{or}\\z=z_{(\posterior,\cdot)}(\mechanismc)\text{ and }\mechanismc\in\mechanismsc^1\cup\mechanismsc^2\end{array}
\\\prior&\text{ if }z=z_\emptyset(\mechanismc)\text{ and }\mechanismc\in\mechanismsc^1\\
\nu_2(\prior,\pi_{\vh}^*(\mechanismc))&\text{ if }z=z_{(\emptyset)}(\mechanismc)\text{ and }\mechanismc\in\mechanismsc^2\\
\nu_3(\prior,\posterior,r_{\vh}^*(\mechanismc,1),r_{\vl}^*(\mechanismc,1))&\text{ if }z=z_{(\posterior,\cdot)}(\mechanismc)\text{ and }\mechanismc\in\mechanismsc^3
\end{array}\right..
\end{align}
Let $\prior$ denote the seller's prior in $\game\prior)$. Define $\mu^*(\emptyset)=\prior$ and $\Gamma^*(\emptyset)=\gamma^*(\prior)$, where $\gamma^*$ is the strategy in the posted-prices assessment \Pbestar. For any public history $h^\dateindex$, define
\begin{align}\label{eq:beliefsfull}
\mu_t^*(h^\dateindex)&=T(\mu_t^*(h^{\dateindex-1}),z(h^\dateindex))).
\end{align}
If either (i) $z=z_{(\posterior,(0,\cdot))}(\mechanismc)$ and $\mechanism\in\mechanismsc^0\cup\mechanismsc^1\cup\mechanismsc^2$, (ii) $z=z_\emptyset(\mechanism)$, $\mu^*(h^\dateindex)\notin\{\delay_i\}_{i\geq 1}$ and $\mechanism\in\mechanismsc^2$, (iii) $z=z_{(\posterior,(0,\cdot))}(\mechanismc)$, $\mu^*(h^\dateindex)\notin\{\delay_i\}_{i\geq 1}$ and $\mechanism\in\mechanismsc^3$, or (iv) $z=z_\emptyset(\mechanism)$ and $\mechanism\in\mechanismsc^0\cup\mechanismsc^1\cup\mechanismsc^3$, define
\begin{align}\label{eq:sellerfull1}
\Gamma_t^*(h^\dateindex)(\mechanism)&=\mathbbm{1}[\mechanism=\gamma^*(T(\mu_t^*(h^{\dateindex-1}),z(h^\dateindex))))].
\end{align}
If $z(h^\dateindex)=z_\emptyset(\mechanism)$ for $\mechanism\in\mechanismsc^2$ and $\mu_t^*(h^\dateindex)=\delay_i,i\geq 1$, define
\begin{align}\label{eq:sellerfull2}
\Gamma_t^*(h^\dateindex)(\mechanism)&=\left\{\begin{array}{cl}\phi_2(\mechanism,\delay_i)&\mechanism=\gamma^*(\delay_i)\\1-\phi_2(\mechanism,\delay_i)
&\mechanism=\gamma^{**}(\delay_{i})\\0&\text{otherwise}\end{array}.\right.
\end{align}
Finally, if $z(h^\dateindex)=z_{(\posterior,(0,\cdot))}(\mechanismc)$ for $\mechanism\in\mechanismsc^3$ and $\mu_t^*(h^\dateindex)=\delay_i,i\geq 1$, define
\begin{align}\label{eq:sellerfull3}
\Gamma_t^*(h^\dateindex)(\mechanism)&=\left\{\begin{array}{cl}\phi_3(\mechanism,\delay_i)&\mechanism=\gamma^*(\delay_i)\\1-\phi_3(\mechanism,\delay_i)
&\mechanism=\gamma^{**}(\delay_{i})\\0&\text{otherwise}\end{array}.\right.
\end{align}
Equations \ref{eq:participationeqbm}-\ref{eq:reportingeqbm}, together with equations \ref{eq:beliefsfull}- \ref{eq:sellerfull3}, define a PBE assessment where the system of beliefs is derived from the strategy profile where possible.

\subsection{Seller's and buyer's sequential rationality}\label{appendix:sellerrationality}
We now check that at all histories, neither the seller nor the buyer have a one-shot deviation from the prescribed strategy profile, given the continuation values constructed in \autoref{appendix:payoffs} and \autoref{appendix:complete}. That this is true for the buyer given the seller's strategy follows from the construction in \autoref{appendix:complete}.

To verify that sequential rationality holds for the seller we proceed in two steps. First, we show that for every \mechanismt\ not in $\mechanismsc^0$, there exists a $\mechanismbt\in\mechanismsc^0$ such that the revenue of offering \mechanismt\ (under the buyer's strategy) coincides with the revenue from offering \mechanismbt. Second, we use this property to relate the seller's payoff from offering \mechanismt\ to the virtual surplus of \mechanismbt\ to show that the revenue from \mechanismt\ is bounded above by $\postedpayoff(\belief^*(\publict))$.

\paragraph{Step 1:} We now show how to obtain \mechanismbt\ from \mechanismt\ for the case in which \mechanismt\ is in $\mechanismsc^2$. The construction for $\mechanismt\in\mechanismsc^1\cup\mechanismsc^3$ follows similar steps. Suppose that the seller with belief $\belief^*(\publict)$ offers a mechanism \mechanismt\ in $\mechanismsc^2$. Let $\pi_{\vh}^*(\hagt,\mechanismt),r_{\vh}^*(\hagt,\mechanismt,1)$ denote the buyer's best responses as constructed in \autoref{appendix:complete}. Denote by $\nu_2\equiv\nu_2(\belief(\publict),\pi_{\vh}^*(\hagt,\mechanism))$  the seller's belief that the buyer's value is \vh\
when he observes non-participation. The seller's payoff is then
\begin{align}\label{eq:payofftype2}
&\prior\pi_{t\vh}^*(\hagt,\mechanismt)\int_ {\Posteriors\times\allocations}\left[\transfer+\delta(1-q)\vh\right]\left(\sum_{v\in V}\varphi^{\mechanismt}(d(\belief,q,\transfer)|v)r_{t\vh}^*(\hagt,\mechanismt)(v)\right)\nonumber\\&+(1-\mu^*(h^\dateindex)\pi_{t\vh}^*(\hagt,\mechanismt))\delta\postedpayoff(\nu_2),
\end{align}
where the continuation values after rejection are constructed using the equilibrium strategy when the seller has posterior $\nu_2$.\footnote{Recall that the seller's continuation payoff only depends on his belief about the buyer's valuation being \vh; it is only the high-valuation buyer's payoff which may be different from \postedpayoffh\ off the path of play.} Now, consider the following alternative mechanism, $\mechanismtprm$:
\small
\begin{align*}
\betacprm(1|\vh)&=\pi_{t\vh}^*(h_B^t,\mechanismt)\\
\betacprm(\nu_2|\vh)&=(1-\pi_{t\vh}^*(h_B^t,\mechanismt)),\quad\betacprm(\nu_2|\vl)=1\\
q^{\mechanismtprm}(\nu_2)&=\transfer^{\mechanismtprm}(\nu_2)=0\\
\transfer^{\mechanismtprm}(1)&=\int_{\Posteriors\times\allocations}\transfer\left(\sum_{v\in V}\varphi^{\mechanismt}(d(\belief,q,\transfer)|v)r_{t\vh}^*(\hagt,\mechanismt)(v)\right)\\
q^{\mechanismtprm}(1)&=\int_{\Posteriors\times\allocations}q\left(\sum_{v\in V}\varphi^{\mechanismt}(d(\belief,q,\transfer)|v)r_{t\vh}^*(\hagt,\mechanismt)(v)\right),
\end{align*}
\normalsize
Note that if the buyer participates and truthfully reports her type,  $\mechanismtprm$ gives the seller a payoff equal to the expression in \autoref{eq:payofftype2}. Then, we can write the payoff to mechanism, $\mechanismt$, as
\begin{align*}
\Payoffs(\publict,\mechanismt)\equiv\sum_{\belief_{t+1}\in\{\nu_2,1\}}\tau^{\mechanismtprm}(\belief_{t+1})[\transfer^{\mechanismtprm}(\belief_{t+1})+\delta(1-q^{\mechanismtprm}(\belief_{t+1}))\postedpayoff(\belief_{t+1})].
\end{align*}

\paragraph{Step 2:} We now show that $\Payoffs(\publict,\mechanismt)$ is bounded above by $\postedpayoff(\belief^*(\publict))$. Indeed, we have:
\begin{align*}
\Payoffs(\publict,\mechanismt)&=\int_{\Posteriors}[\transfer^{\mechanismbt}(\belief_{t+1})+(1-q^{\mechanismbt}(\belief_{t+1}))\delta\postedpayoff(\belief_{t+1})]\tau^{\mechanismbt}(d\belief_{t+1})\\
&\leq\virtual(\tau^{\mechanismbt},q^{\mechanismbt},(\payoffh^{\mechanism},\postedpayoff),\belief(\publict))\\
&\leq\virtual(\tau^{\mechanismbt},q^{\mechanismbt},(\postedpayoffh,\postedpayoff)_{\belief_{t+1}<\belief(\publict)},(\payoffh^{\mechanism},\postedpayoff)_{\belief_{t+1}\geq\belief(\publict)}),\belief(\publict))\\
&\leq\max_{(\tau,q),(\payoffh,\payoffs)_{\belief_{t+1}>\belief(\publict)})}\virtual(\tau^{\mechanismbt},q^{\mechanismbt},(\postedpayoffh,\postedpayoff)_{\belief_{t+1}<\belief(\publict)},(\payoffh,\payoffs)_{\belief_{t+1}\geq\belief(\publict)}),\belief(\publict))=\postedpayoff(\belief(\publict))
\end{align*}
The first equality represents the payoff from offering \mechanismt\ as the payoff from offering  the mechanism \mechanismbt\ that satisfies the incentive compatibility and participation constraints (note that the seller's continuation payoffs are those of the assessment \Pbestar\ even after a deviation). The first inequality follows from \autoref{lemma:vs-upper bound}. Now, while the seller's continuation payoffs are given by \postedpayoff, the buyer's continuation payoffs need not be exactly those of \Pbestar\ when the seller's posterior belief is one of the threshold beliefs. The second inequality follows from noting that when $\belief_{t+1}<\belief(\publict)$, the term multiplying \payoffh\ in the virtual surplus is negative. Thus, the seller with belief \belief(\publict) prefers that whenever his posterior belief $\belief_{t+1}$ is below \belief(\publict), the high-valuation buyer's continuation payoff is that which minimizes her rents, i.e., $\postedpayoffh(\belief_{t+1})$. The last inequality follows by definition. The proofs in Sections \ref{sec:opt}, \ref{sec:uniqueness}, and \ref{appendix:opt} imply the last equality. We conclude that the seller has no one-shot deviations. 
\section{Perfect Bayesian equilibrium: formal statement}\label{appendix:pbe}
We introduce in this section the necessary formalisms to define PBE. To simplify notation, we assume in what follows that \mechanisms\ is such that $M_i$ is finite for all $i\in\pazocal{I}$. When \Types\ is finite, it follows from \cite{doval2020mechanism} that this is without loss of generality. It affords two important simplifications. First, \mechanisms\ is itself a Polish space, which means that we can condition the buyer's strategy directly on the mechanism, \mechanismc, chosen by the seller, as opposed to the realization of the randomization device in the construction in \cite{aumann1964mixed}. Second, given a public history $h^t$, the set of buyer histories consistent with $h^t$, $H_B^t(h^t)$, is finite and therefore the support of $\mu_t(h^t)\in\Delta(\Types\times H_B^t(h^t))$ is finite. 
\newline\indent
To define the seller and the buyer's payoffs, we introduce the following notation. Given a period $t$ and a mechanism \mechanismt, we let \zpt\ denote the tuple $(\mechanismt,1,s_t,(q_t,\transfer_t))$ and \znpt\ denote the tuple $(\mechanismt,0,\emptyset, (0,0))$. Furthermore, given the buyer's participation and reporting strategy and a mechanism \mechanismt\ define a new distribution over $\Delta(\St\times\{0,1\}\times\mathbb{R})$ so that
\begin{align*}
\rho^{(\pi,r)}(S^\prime\times\allocations^\prime|v,h_B^t,\mechanismt)=\pi_{tv}(h_B^t,\mechanismt)\sum_{m\in\Mt}\varphi^{\mechanismt}(S^\prime\times\allocations^\prime|m)r_{tv}(h_B^t,\mechanismt,1)(m)
\end{align*}
for all measurable subsets $S^\prime\times A^\prime\subseteq\St\times\{0,1\}\times\mathbb{R}$.
\newline\indent
Fix an assessment \Pbe, a public history $h^t$ and a mechanism, \mechanismt. The seller's payoffs are given by
\small
\begin{align*}
&U_S(\Gamma,(\pi_v,r_v)_{v\in\Types},\mu|h^t,\mechanismt)=\sum_{(v,h_B^t)}\mu_t(h^t)(v,h_B^t)(1-\pi_{tv}(h_B^t,\mechanism_t))\delta \mathbb{E}_{\Gamma}\left[U_S(\Gamma,(\pi_v,r_v)_{v\in\Types},\mu|h^t,\znpt,\cdot)\right]\\
&+\sum_{(v,h_B^t)}\mu_t(h^t)(v,h_B^t)\int_{\St\times\{0,1\}\times\mathbb{R}}\left(\transfer+(1-q)\delta\mathbb{E}_\Gamma\left[ U_S(\Gamma,(\pi_v,r_v)_{v\in\Types},\mu|h^t,z_{(s_t,(0,\transfer))}(\mechanismt),\cdot)\right]\right)\rho^{(\pi,r)}(d(s_t,q,\transfer)|v,h_B^t,\mechanismt).
\end{align*}
\normalsize
Similarly, the buyer's payoffs when her valuation is $v$, the history is $h_B^t$, and the seller offers mechanism \mechanismt\ are given by
\small
\begin{align*}
&U_v(\Gamma,(\pi_v,r_v),\mu|h_B^t,\mechanismt)=(1-\pi_{tv}(h_B^t,\mechanismt))\delta\mathbb{E}_{\Gamma}\left[U_v(\Gamma,(\pi_v,r_v),\mu|h_B^t,\znpt,\cdot)\right]+\pi_{tv}(h_B^t,\mechanismt)\times
\\&
\sum_{m\in\Mt}r_{tv}(h_B^t,\mechanismt,1)(m)\int_{\St\times\allocations}\left(vq-\transfer+(1-q)\delta\mathbb{E}_\Gamma\left[U_v(\Gamma,(\pi_v,r_v)_{v\in V},\mu|h_B^t,m,z_{(s_t,(q,x))}(\mechanismt),\cdot)\right]\right)\varphi^{\mechanismt}(d(s_t,q,x)|m)
\end{align*}
\normalsize
\begin{definition}\label{definition:pbe-sr} The assessment \Pbe\ satisfies sequential rationality if for all periods $t$, and all public histories $h^t$, we have
\begin{enumerate}
\item For all mechanisms \mechanismt\ in the support of \space$\Gamma_t(h^t)$, $U_S(\Gamma,(\pi_v,r_v)_{v\in\Types},\mu|h^t,\mechanismt)\geq\\ U_S(\Gamma,(\pi_v,r_v)_{v\in\Types},\mu|h^t,\mechanismt^\prime)$ for all $\mechanismt^\prime\neq\mechanismt$,
\item For all $v\in V$, all buyer histories $h_B^t\in H_B^t(h^t)$, and all mechanisms \mechanismt, $U_v(\Gamma,(\pi_v,r_v),\mu|h_B^t,\mechanismt)\geq U_v(\Gamma,(\pi_v^\prime,r_v^\prime),\mu|h_B^t,\mechanismt)$ for all alternative strategies $(\pi_v^\prime,r_v^\prime)$.
\end{enumerate}
\end{definition}
\begin{definition}\label{definition:brwp}
The belief system $\mu$ satisfies Bayes' rule where possible if for all public histories $h^t$, and mechanisms $\mechanismt$, the following hold,
\small
\begin{align*}
&\mu_{t+1}(h^t,\znpt)(\vh,h_B^t,\znpt)\left[\sum_{v\in\Types,h_B^t\in H_B^t(h^t)}\mu_t(h^t)(v,h_B^t)(1-\pi_{tv}(h_B^t,\mechanismt))\right]=\mu_t(h^t)(\vh,h_B^t)(1-\pi_{t\vh}(h_B^t,\mechanismt)),
\end{align*}\normalsize
and, for all measurable subsets $S^\prime\times A^\prime$ of $\St\times\allocations$
\small\begin{align*}
&\sum_{(v,h_B^t)}\mu_t(h^t)(v,h_B^t)\int_{S^\prime\times\allocations^\prime}\mu_{t+1}(h^t,\cdot)(\overline{v},\overline{h}_B^t,\zpt,\overline{m})\rho^{(\pi,r)}(d(s_t,q,\transfer)|v,h_B^t)\\
&
=\mu_t(h^t)(\overline{v},\overline{h}_B^t)\pi_{t\overline{v}}(\overline{h}_B^t,\mechanismt)r_{t\overline{v}}(\overline{h}_B^t,\mechanismt,1)(\overline{m})\varphi^{\mechanismt}(S^\prime\times A^\prime|\overline{m}).
\end{align*}
\normalsize
\end{definition}
\begin{definition}\label{definition:pbe}
An assessment \Pbe\ is a Perfect Bayesian equilibrium if it is sequentially rational and satisfies Bayes' rule where possible.
\end{definition}
\end{document}